\documentclass{siamonline0316}

\usepackage[latin1]{inputenc}
\usepackage[T1]{fontenc}
\usepackage{xspace}
\usepackage[shortlabels]{enumitem}
\usepackage[font=footnotesize]{caption}  

\usepackage[]{cite} 
\usepackage[numbers]{natbib}  
\usepackage[labeled,resetlabels]{multibib}
\newcites{SM}{References for the Supplementary Materials}

\usepackage{amsmath,amsfonts,amssymb,bm}
\usepackage{dsfont}             
\renewcommand{\mathbb}{\mathds} 
\usepackage[centercolon = true]{mathtools}

\usepackage{graphicx}
\usepackage[captionskip=4mm]{subfig}
\usepackage[centering]{geometry}

\theorembodyfont{\normalfont}
\newtheorem{remark}{Remark}

\newcommand \SetTableFontSize \footnotesize

\newlength{\dhatheight}
\newcommand{\doublehat}[2]{%
    \settoheight{\dhatheight}{\ensuremath{\hat{#1}}}%
    \addtolength{\dhatheight}{-0.35ex}%
    \mskip #2\hat{\vphantom{\rule{1pt}{\dhatheight}}%
    \smash{\mskip -#2\hat{#1}}}}

\newcommand   \RR     {\mathbb{R}}
\newcommand   \NN     {\mathbb{N}}
\newcommand   \XX     {\mathbb{X}}
\newcommand   \YY[1]  {\mathbb{Y}_{#1}} 
\newcommand   \Y[2]   {Y_{#1}^{#2}}     
\newcommand   \Yt[2]  {\widetilde{Y}_{#1}^{#2}} 
\renewcommand \P      {\mathsf{P}}   
\newcommand   \PX     {\P_{\XX}}     
\newcommand   \ddpX   {\pi_{\XX}}    
\newcommand   \E      {\mathsf{E}}   
\newcommand   \EE     {\E}           
\newcommand   \Bcal   {\mathcal{B}}  
\newcommand   \Ncal   {\mathcal{N}}  
\newcommand   \alphaMC  {\hat{\alpha}^{\mskip 1mu \text{\tiny MC}}}
\newcommand   \alphaSS  {\hat{\alpha}^{\mskip 1mu \text{\tiny SS}}}
\newcommand   \alphaB   {\hat{\alpha}^{\mskip 1mu \text{\tiny B}}}
\newcommand   \alphaBSS {\doublehat{\alpha}{1mu}^{\mskip 1mu \text{\tiny BSS}}}
\newcommand   \pB       {\hat{p}^{\mskip 3mu \text{\tiny B}}}
\newcommand   \pBSS     {\doublehat{p}{3mu}^{\mskip 3mu \text{\tiny BSS}}}
\newcommand   \g        {\tilde g}        
\renewcommand \u[1]     {u_{#1}}          
\newcommand   \one    {{\mathbb{1}}}

\newcommand   \dotvar {\,\bm{\cdot}\,}
\newcommand   \ddiff  {\mathrm{d}}

\newcommand   \dx     {\ddiff x}
\newcommand   \dy     {\ddiff y}
\newcommand   \dPX    {\ddiff\PX}
\DeclareMathOperator \var  {var}

\DeclareMathOperator \card  {card}
\DeclareMathOperator* \argmin {argmin}

\renewcommand{\hat}{\widehat}

\newcommand{\bmid}{\;\big\lvert\;}
\newcommand \symdiff {\mathrel{\triangle}}
\newcommand \TwoSMART {${}^2$SMART\xspace}
\newcommand \xihat {\hat{\xi}}
\newcommand \iid {IID\xspace}
\newcommand \thres {u}

\newcommand \eqdef {\mathrel{\coloneqq}}  

\newcommand \TheTitle {Bayesian subset simulation}
\newcommand \TheAuthors {Julien Bect, Ling Li and Emmanuel Vazquez}

\headers{\TheTitle}{\TheAuthors}

\title{{\TheTitle}\thanks{%
    This research was partially funded by the French \emph{Fond Unique
      Interministériel} (FUI 7) in the context of the \textsc{CSDL}
    (Complex Systems Design Lab) project.  Parts of this work were
    previously published in the proceedings of the PSAM~11 \& ESREL~12
    conference~\cite{lilyPSAM} and in the PhD thesis of the second
    author~\cite{lily:phd}.}}

\author{%
  Julien Bect%
  \thanks{%
    Laboratoire des Signaux et Systèmes,
    CentraleSupélec, CNRS, Univ. Paris-Sud, Université Paris-Saclay.
    3~rue Joliot-Curie, 92192 Gif-sur-Yvette, France. %
    Email: \email{firstname.lastname@centralesupelec.fr}%
  }%
  \and%
  Ling Li%
  \thanks{%
    Email: \email{ling.li.supelec@gmail.com}. %
    \emph{Present address:} Schlumberger Gould Research Center, Cambridge, UK.%
  }
  \and%
  Emmanuel Vazquez\footnotemark[2]
}

\hypersetup{
  pdftitle={\TheTitle},
  pdfauthor={\TheAuthors},
  bookmarksnumbered,
  citecolor=blue
}

\begin{document} 

\maketitle

\begin{abstract}
  We consider the problem of estimating a probability of failure~$\alpha$,
  defined as the volume of the excursion set of a function
  $f:\mathbb{X} \subseteq \mathbb{R}^{d} \to \mathbb{R}$ above a given
  threshold, under a given probability measure on~$\mathbb{X}$. In this article,
  we combine the popular subset simulation algorithm (Au and Beck,
  Probab.~Eng.~Mech.~2001) and our sequential Bayesian approach for the
  estimation of a probability of failure (Bect, Ginsbourger, Li, Picheny and
  Vazquez, Stat. Comput. 2012). This makes it possible to estimate~$\alpha$ when
  the number of evaluations of~$f$ is very limited and~$\alpha$ is very
  small. The resulting algorithm is called Bayesian subset simulation (BSS).  A
  key idea, as in the subset simulation algorithm, is to estimate the
  probabilities of a sequence of excursion sets of~$f$ above intermediate
  thresholds, using a sequential Monte Carlo (SMC) approach.  A Gaussian process
  prior on~$f$ is used to define the sequence of densities targeted by the SMC
  algorithm, and drive the selection of evaluation points of~$f$ to estimate the
  intermediate probabilities.  Adaptive procedures are proposed to determine the
  intermediate thresholds and the number of evaluations to be carried out at
  each stage of the algorithm.  Numerical experiments illustrate that BSS
  achieves significant savings in the number of function evaluations with
  respect to other Monte~Carlo approaches.
\end{abstract}

\begin{keywords}
  Probability of failure, %
  Computer experiments, %
  Sequential design, %
  Gaussian process, %
  Stepwise uncertainty reduction %
  Sequential Monte Carlo
\end{keywords}

\begin{AMS}
  62L05, 
  62K99, 
  62P30  
\end{AMS}

\section{Introduction}
\label{sec:intro}

Probabilistic reliability analysis has become over the last thirty years an
essential part of the engineer's toolbox (see, e.g., \cite{melchers:1999:book,
  deroc:2008:book, oconnor:2012:book}). One of the central problems in
probabilistic reliability analysis is the computation of the probability of
failure
\begin{equation}
  \label{eq:def-alpha}
  \alpha \;=\; \int_\XX \one_{f \le 0}\, \dPX
\end{equation}
of a system (or a component in a multicomponent system; see, e.g.,
\cite{rausand:2004:book}, where $\PX$ is a probability measure over some
measurable space~$\left( \XX, \Bcal \right)$ representing all possible sources
of uncertainty acting on the system---both epistemic and aleatory---and
$f:\XX \to \RR$ is the so-called \emph{limit-state function}, such that $f$
takes positive values when the system behaves reliably, and negative values when
the system behaves unreliably, or fails. It is assumed in this article
that~$\XX$ is a subset of~$\RR^d$---in other words, we consider reliability
problems where all uncertain factors can be described as a $d$-dimensional
random vector. Numerous examples of applications that fall into this category
can be found in the literature (see, for instance, \cite{waarts:2000:phd,
  jonkman:2008:flood, deroc:2008:chap5, bayarri:2009:volcanic, zio:2009:loca,
  garcia:2010:emc}).

Two major difficulties usually preclude a brute-force Monte Carlo (MC) approach,
that is, using the estimator
\begin{equation*}
  \alphaMC = \frac{1}{m}\sum_{i=1}^m \one_{f(X_i) \le 0}\,,\quad
  X_i\stackrel{\rm i.i.d}{\sim} \PX\,,
\end{equation*}
which requires $m$ evaluations of~$f$.  First, the evaluation of $f$ for a
given~$x \in \XX$ often relies on one or several complex computer programs
(e.g., partial differential equation solvers) that take a long time to
run. Second, in many applications, the failure
region~$\Gamma = \{ x \in \XX \mid f(x) \le 0 \}$ is a \emph{rare event} under
the probability~$\PX$; that is, the probability of failure $\alpha=\PX(\Gamma)$
is small. When $\alpha$ is small, the standard deviation of $\alphaMC$ is
approximately $\sqrt{\alpha/m}$.  To estimate~$\alpha$ by MC with a standard
deviation of~$0.1\alpha$ thus requires approximately $100/\alpha$ evaluations of
$f$. As an example, with~$\alpha = 10^{-3}$ and~$10$ minutes per evaluation,
this means almost two years of computation time.

The first issue---designing efficient algorithms to estimate~$\alpha$ in the
case of an expensive-to-evaluate limit-state function---can be seen as a problem
of \emph{design and analysis of computer experiments} (see, e.g.,
\cite{santner:2003:dace}), bearing some similarities to the problem of global
optimization (see \cite{villemonteix:2009:iago} and references therein). Several
sequential design strategies based on Gaussian process models have been proposed
in the literature, and spectacular evaluation savings have been demonstrated on
various examples with moderately small~$\alpha$ (typically, $10^{-2}$ or
$10^{-3}$); see \cite{bect:2012:stco} for a review of fully sequential
strategies and \cite{dubourg:icasp11, auffray2014rare} for examples of two-stage
strategies. The closely related problem of quantile estimation has also been
investigated along similar lines \cite{oakley:2004:perc, cannamela:2008,
  arnaud:2010:quant}.

A key idea to address the second issue---i.e., to estimate a small probability
of failure---is to consider a decreasing sequence of events
$\Gamma_1 \supset \Gamma_2 \supset \cdots \supset \Gamma_T = \Gamma$ such that
the conditional probabilities $\PX\left( \Gamma_t \mid \Gamma_{t - 1} \right)$
are reasonably large, and therefore easier to estimate than~$\alpha$
itself. Then, sequential Monte Carlo simulations \cite{delmoral:2006:sequential}
can be used to produce estimates~$\hat p_t$ of the conditional probabilities
$\PX\left( \Gamma_t \mid \Gamma_{t - 1} \right)$, leading to a product-form
estimate $\prod_{t = 1}^T \hat p_t$ for~$\alpha$. This idea, called \emph{subset
  simulation}, was first proposed in \cite{au01:_estim} for the simulation of
rare events in structural reliability analysis\footnote{A very similar algorithm
  had in fact been proposed earlier by \cite{diaconis:1995:three}, but for a
  quite different purpose (estimating the probability of a rare event under the
  bootstrap distribution).}, but actually goes back to the much older
\emph{importance splitting} (or \emph{multilevel splitting}) technique used for
the simulation of rare events in Markovian models (see, e.g.,
\cite{glasserman:1999:mls} and references therein). Subset simulation has since
then become one of the most popular techniques for the computation of small
probabilities of failure, and the theoretical properties of several (most of the
times idealized) variants of the algorithm have recently been investigated by
several authors (see, e.g., \cite{cerou2012, brehier2016}). However, because of
the direct use of a Monte Carlo estimator for~$\hat p_t$ at each stage~$t$, the
subset simulation algorithm is not applicable when~$f$ is expensive to evaluate.

In this article we propose a new algorithm, called \emph{Bayesian subset
  simulation} (BSS), which tackles both issues at once using ideas from the
sequential design of computer experiments and from the literature on sequential
Monte Carlo methods.  Section~\ref{sec:subsim} reviews the subset simulation
algorithm from the point of view of sequential Monte Carlo (SMC) techniques to
prepare the ground for the introduction of our new
algorithm. Section~\ref{sec:bss} describes the algorithm itself and
Section~\ref{sec:num} presents numerical results. Finally,
Section~\ref{sec:discuss} concludes the article with a discussion.

\section{Subset simulation: a sequential Monte Carlo algorithm}
\label{sec:subsim}

This section recalls the main ideas of the classical subset simulation algorithm
\cite{au01:_estim}, which, although not originally presented as such, can be
seen as a sequential Monte Carlo sampler \cite{delmoral:2006:sequential,
  cerou2012}.

\subsection{Idealized subset simulation (with fixed levels and \iid sampling)}
\label{sec:ideal-subs-simul}

We consider the problem of estimating the probability~$\alpha$ of a rare
event~$\Gamma$ of the form $\Gamma = \left\{ x \in \XX : f(x) > u \right\}$,
where $u \in \RR$ and $f:\XX \to \RR$, using pointwise evaluations of~$f$. Note
that the limit-state function (see Section~\ref{sec:intro}) can be defined as
$x \mapsto u - f(x)$ with our notations. Assuming, for the sake of simplicity,
that~$\PX$ has a probability density function~$\ddpX$ with respect to Lebesgue's
measure, we have
\begin{equation*}
  \alpha = \int_\XX \one_{f(x) > u}\, \ddpX (x)\, \dx \,.
\end{equation*}

The key idea of the subset simulation algorithm is to introduce an increasing
(finite) sequence of thresholds $-\infty = u_0 < u_1 < u_2 \cdots < u_T = u$,
which determine a corresponding decreasing sequence of subsets:
\begin{equation*}
  \XX = \Gamma_0 \supset \Gamma_1 \supset \cdots \supset \Gamma_T = \Gamma,
  \quad \Gamma_t \eqdef \left\{ x \in \XX : f(x) > u_t \right\},
\end{equation*}
of the input space~$\XX$. Let $\alpha_t = \PX\left( \Gamma_t \right)$. The
decreasing sequence~$\left( \alpha_t \right)_{0 \le t \le T}$ obeys the
recurrence formula
\begin{equation}
  \label{eq:recurr-SS}
  \alpha_{t + 1}
  = \alpha_t\, \PX\left( \Gamma_{t + 1} \mid \Gamma_t \right)
  = \alpha_t\, \int \one_{\Gamma_{t + 1}}(x)\, q_t(x)\, \dx,
\end{equation}
where $q_t$ stands for the truncated density 
\begin{equation}
  \label{equ:truncated-density}
  q_t(x) \;=\; \frac{\one_{\Gamma_t}(x)\, \ddpX(x)}{%
    \int \one_{\Gamma_t}(y)\, \ddpX(y)\, \dy} \,.
\end{equation}
The small probability~$\alpha = \alpha_T$ can thus be rewritten as a product of
conditional probabilities, which are larger (and therefore easier to estimate)
than~$\alpha$:
\begin{equation*}
  \alpha = \prod_{t=1}^T\, p_t, \quad
  p_t \eqdef \PX\left( \Gamma_{t} \mid \Gamma_{t - 1} \right).
\end{equation*}

Assume that, for each~$t \in \{ 0, 1, \ldots, T - 1\}$, a
sample~$\bigl( Y_t^j \bigr)_{1 \le j \le m}$ of independent and identically
distributed (\iid) random variables from the truncated density~$q_t$ is
available. Then, each conditional probability~$p_t$ can be estimated by the
corresponding Monte-Carlo estimator
$\hat p_t = \frac{1}{m} \sum_{j = 1}^m \one_{\Gamma_t} \bigl( Y_{t-1}^j \bigr)$,
and $\alpha$ can be estimated by the product-form estimator
$\alphaSS = \prod_{t = 1}^T \hat p_t$. By choosing the thresholds $u_t$ in such
a way that the conditional probabilities $p_{t}$ are high, $\alpha$ can be
estimated using fewer evaluations of~$f$ than what would have been necessary
using a simple Monte Carlo approach (see Section~\ref{sec:subsim-adapt-thresh}
for a quantitative example).

\subsection{Sequential Monte-Carlo simulation techniques}
\label{sec:sequ-monte-carlo}

Generating exact \iid draws from the densities~$q_t$ is usually not possible, at
least not efficiently, even if a method to generate \iid samples
from~$q_0 = \ddpX$ is available. Indeed, although the accept-reject algorithm
(see, e.g., \cite{robert:2004:monte}, Section~2.3) could be used in principle,
it would be extremely inefficient when~$t$ is close to~$T$, that is,
when~$\PX\left\{ \Gamma_t \right\}$ becomes small. This is where sequential
Monte-Carlo (SMC) simulation techniques are useful.

Given a sequence~$\left( q_t \right)_{0 \le t < T}$ of probability density
functions over~$\XX$, SMC samplers sequentially generate, for each target
density $q_t$, a weighted
sample~$\YY{t} = \bigl( \bigl( w_t^j, Y_t^j \bigr) \bigr)_{1 \le j \le m}$,
where $w_t^j \ge 0$, $\sum_j w_t^j = 1$ and $Y_t^j \in \XX$. The random vectors
$Y_t^j$ are usually called \emph{particles} in the SMC literature, and the
weighted sample~$\YY{t}$ is said to \emph{target} the distribution~$q_t$. They
are, in general, neither independent nor distributed according to~$q_t$, but
when the sample size~$m$ goes to infinity, their empirical distribution
$\mu_t^{(m)} = \sum_{j=1}^m w_t^{j}\, \delta_{Y_t^j}$ converges to the target
distribution---that~is, to the distribution with probability density
function~$q_t$---in the sense that
\begin{equation*}
  \int_\XX h(x)\, \ddiff\mu_t^{(m)}(x)
  \;=\; \sum_{j=1}^m w_t^j\, h(Y_t^j)
  \;\to\; \int_\XX h(x)\, q_t(x)\, \dx,
\end{equation*}
for a certain class of integrable functions~$h$.

In practice, each weighted sample~$\YY{t}$ is generated from the previous one,
$\YY{t-1}$, using transformations; SMC algorithms are thus expected to be
efficient when each density~$q_t$ is, in some sense, close to its predecessor
density~$q_{t - 1}$. The specific transformations that are used in the subset
simulation algorithm are described next. The reader is referred to
\cite{delmoral:2006:sequential, liu:2008:book} and references therein for a
broader view of SMC sampling techniques, and to \cite{douc:2008:limit} for some
theoretical results on the convergence (law of large numbers, central limit
theorems) of SMC algorithms.

\subsection{Reweight/resample/move}
\label{sec:reweight--resample}

We now describe the reweight/resample/\allowbreak move scheme that is used in
the subset simulation algorithm to turn a weighted sample $\YY{t-1}$ targeting
$q_{t-1} \propto \one_{\Gamma_{t-1}}\, \ddpX$ into a weighted sample $\YY{t}$
targeting $q_t \propto \one_{\Gamma_t}\, \ddpX$. This scheme, used for instance
in \cite{chopin:2002}, can be seen as a special case of the more general SMC
sampler of~\cite{delmoral:2006:sequential}\footnote{See in particular
  Section~3.1.1, Remark~1, and Section~3.3.2.3.}.

Assume a weighted sample $\YY{t-1} = \bigl( \bigl( w_{t-1}^j, Y_{t-1}^j \bigr)
\bigr)_{1 \le j \le m}$ targeting $q_{t-1}$ has been obtained at stage $t-1$.
The \emph{reweight} step produces a new weighted sample $\YY{t, 0} = \bigl(
\bigl( w_{t,0}^j, Y_{t-1}^j \bigr) \bigr)_{1 \le j \le m}$ that targets~$q_t$,
by changing only the weights in~$\YY{t-1}$:
\begin{equation*}
  w_{t,0}^j \;\propto\; \frac{
    q_t\bigl(Y_{t-1}^j\bigr)
  }{
    q_{t-1}\bigl(Y_{t-1}^j\bigr)
  }\; w_{t-1}^j.
\end{equation*}

The \emph{resample} and \emph{move} steps follow the reweighting step. These
steps aim at avoiding the degeneracy of the sequence of weighted samples---i.e.,
the accumulation of most of the probability mass on a small number of particles
with large weights.

The simplest variant of resampling is the \emph{multinomial resampling}
scheme. It produces a new weighted sample~$\YY{t, 1} = \bigl( \bigl( w_t^j,
Y_{t,1}^j \bigr) \bigr)_{1 \le j \le m}$, where the new particles $Y_{t,1}^j$
have equal weights $w_t^j = \frac{1}{m}$, and are independent and identically
distributed according to the empirical distribution $\sum_{j=1}^m w_{t,0}^{j}\,
\delta_{Y_{t-1}^j}$. In this work, we use the slightly more elaborate
\emph{residual resampling} scheme (see, e.g., \cite{liu:2008:book}), which is
known to outperform multinomial resampling
(\cite{douc2005comparison}, Section~3.2). As in multinomial resampling, the
residual resampling scheme produces a weighted sample with equal weights
$w_t^j=\frac{1}{m}$.

The resampling step alone does not prevent degeneracy, since the resulting
sample contains copies of the same particles. The move step restores some
diversity by moving the particles according to a Markov transition kernel~$K_t$
that leaves~$q_t$ invariant:
\begin{equation*}
  \int q_t(x)\, K_t (x, \dx') \;=\; q_t(x')\, \dx';
\end{equation*}
for instance, a random-walk Metropolis-Hastings (MH) kernel (see,
e.g., \cite{robert:2004:monte}).

\begin{remark}
  In the special case of the subset simulation algorithm, all weights are
  actually equal \emph{before} the reweighting step and, considering the
  inclusion~$\Gamma_t \subset \Gamma_{t - 1}$, the reweighting formula takes the
  form
  \begin{equation*}
    w_{t,0}^j \;\propto\; \one_{\Gamma_t}(Y_{t-1}^j).
  \end{equation*}
  In other words, the particles that are outside the new subset~$\Gamma_t$ are
  given a zero weight, and the other weights are simply normalized to sum to
  one. Note also that the resampling step discards particles outside
  of~$\Gamma_t$ (those with zero weight at the reweighting step).
\end{remark}

\begin{remark}
  Note that Au and Beck's original algorithm \citep{au01:_estim} does
  not use separate resample/move steps as described in this section. %
  Instead, it uses a slightly different (but essentially
  similar) sampling scheme to populate each level: assuming
  that~$L_t = m / m_t$ is an integer, where $m_t$ denote the number of
  particles from stage~$t-1$ that belong to~$\Gamma_t$, they start
  $m_t$ independent Markov chains of length~$L_t$ from each of
  the particles (called ``seeds''). %
  Both variants of the algorithm have the property, in the case of
  fixed levels, that the particles produced at level~$t$ are exactly
  distributed according to~$q_t$.  
\end{remark} 

\begin{remark}
  In the general version of the reweight/resample/move procedure,
  the resampling step is carried out only
  when some degeneracy criterion---such as the expected sample size
  (ESS)---falls below a threshold (see,
  e.g., \cite{delmoral:2006:sequential, DelMoral2012}).
\end{remark}

\subsection{Practical subset simulation: adaptive thresholds}
\label{sec:subsim-adapt-thresh}

It is easy to prove that the subset simulation
estimator~$\alphaSS = \prod_{t = 1}^T \hat p_t$ is unbiased. Moreover, according
to~Proposition~3 in~\cite{cerou2012}, it is asymptotically normal in the
large-sample-size limit:
\begin{equation}
  \label{eq:clt}
  \sqrt{m}\; \frac{\alphaSS - \alpha}{\alpha}
  \;\xrightarrow[m \to \infty]{\mathcal{D}}\;
  \mathcal{N} \left(0; \sigma^2 \right)\,
\end{equation}
where $\xrightarrow{\mathcal{D}}$ denotes convergence in distribution and
\begin{equation}
  \label{eq:sigma2-subsim}
  \sigma^2 \approx \sum_{t=1}^T  \frac{1 - p_t}{p_t}\,,
\end{equation}
when the MCMC kernel has good mixing properties (see \cite{cerou2012}
article for the exact expression of~$\sigma^2$). For a given number~$T$ of
stages, the right-hand side of~\eqref{eq:sigma2-subsim} is minimal when all
conditional probabilities are equal; that is, when $p_t = \alpha^{1/T}$.

In practice however, the value of~$\alpha$ is of course unknown, and it is not
possible to choose the sequence of threshold beforehand in order to make all the
conditional probabilities equal. Instead, a value~$p_0$ is chosen---say, $p_0 =
10\%$---and the thresholds are tuned in such a way that, at each stage~$t$,
$\hat p_t = p_0$. A summary of the resulting algorithm is provided in
Table~\ref{tab:subsim_algo}.

Equations~\eqref{eq:clt} and~\eqref{eq:sigma2-subsim} can be used to quantify
the number of evaluations of~$f$ required to reach a given coefficient of
variation with the subset simulation estimator~$\alphaSS$. Indeed, in the case
where all conditional probabilities are equal, we have
\begin{equation}
  \label{eq:approx-variance}
  \var\left( \alphaSS/\alpha \right) \approx \frac{T}{m}\, \frac{1 - p_0}{p_0}.
\end{equation}
with $T = \log(\alpha) / \log(p_0)$.  For example, take $\alpha = 10^{-6}$. With
the simple Monte Carlo estimator, the number of evaluations of~$f$ is equal to
the sample size~$m$: approximately $n=\delta^{-2}\, \alpha^{-1} = 10^8$
evaluations are required to achieve a coefficient of
variation~$\delta = \mathop{\rm std}({\alphaMC})/\alpha = 10\%$. In contrast,
with $p_0 = 10\%$, the subset simulation algorithm will complete
in~$T = \log(\alpha) / \log(p_0) = 6$ stages, thus achieving a coefficient of
variation $\delta = \mathop{\rm std}({\alphaSS})/\alpha = 10\%$ with
$m = \delta^{-2}\, T\, (1-p_0)/p_0 = 5400$ particles. Assuming that the move
step uses only one evaluation of~$f$ per particle, the corresponding number of
evaluations would be $n = m + (T-1)(1-p_0)m = 29700 \ll 10^8$.

\begin{remark} \label{rem:choix-p0}
  The value $p_0 = 0.1$ was used in the original paper of Au and Beck, on the
  ground that it had been ``found to yield good efficiency''
  \citep[see][Section~5]{au01:_estim}.
  Based on the approximate variance formula~\eqref{eq:approx-variance}, Zuev and
  co-authors \cite{Zuev2012} argue that the variance is roughly proportional for
  a given total number of evaluations to
  $(1 - p_0) / \left( p_0\, (\log (p_0))^2 \right)$, and conclude%
  \footnote{%
    Their analysis is based on the observation that the total number of
    evaluations is equal to~$m\, T$--- in other words, that~$m$ new samples must
    be produced at each stage.
    Some authors \citep[e.g.,][]{brehier2016} consider a variant where the
    particles that come from the previous stage are simply copied to the new set
    of particles, untouched by the Move step.
    In this case, a similar analysis suggests that 1) the optimal value of~$p_0$
    actually depends on~$\alpha$, and is somewhere between~0.63 (for
    $\alpha = 0.01$) and~1.0 (when $\alpha \to 0$); and 2) the value
    of~$\delta^2$ is only weakly dependent on~$p_0$, as long as~$p_0$ is not too
    close to~$0$ (say, $p_0 \ge 0.1$).
  } %
  that any $p_0 \in \left[ 0.1; 0.3 \right]$ should yield
  quasi-optimal results, for any~$\alpha$.
\end{remark}

\begingroup
\linespread{0}

\begin{table} 
  \SetTableFontSize

  \caption{Subset simulation algorithm with adaptive thresholds}
  \label{tab:subsim_algo}

  \noindent\rule{\textwidth}{0.4pt}
  \begin{enumerate}[itemsep=2ex]

  \item[] Prescribe $m_0<m$ a fixed number of ``succeeding particles''. Set
    $p_0= \frac{m_0}{m}.$
    
  \item {\bf Initialization } (stage $0$) 
    \begin{enumerate}[topsep=0pt, itemsep=0pt]
    \item Generate an $m$-sample $Y_0^j \stackrel{\rm i.i.d}{\sim} \P_{\XX}$,
      $1\leq j \leq m$, and evaluate $f\left( Y_0^j \right)$ for all~$j$.
    \item Set $u_0 = -\infty$ and~$t = 1$.
    \end{enumerate}
  
  \item {\bf Repeat } (stage~$t$) 
    \begin{enumerate}

    \item \textbf{Threshold adaptation}
      \begin{itemize}[topsep=0pt, itemsep=0pt]
      \item Compute the $(m-m_0)$-th order statistic of
        $\bigl( f(Y_{t-1}^{j}) \bigr)_{1 \le j \le m}$ and call it~$u_t^0$.
      \item If $u_t^0 > u$, set $u_t = u$, $T = t$ and go to the estimation
        step.  \vspace{1mm}
      \item Otherwise, set $u_t = u_t^0$ and
        $\Gamma_t = \{x\in\XX; f(x) > u_t \}$.
      \end{itemize}

    \item \textbf{Sampling}
      \begin{itemize}[topsep=0pt, itemsep=-1mm]
      \item \emph{Reweight}: set
        $m_t = \card \{j \le m \colon Y^j_{t-1} \in \Gamma_t \}$ and
        $w_{t,0}^j = \frac{1}{m_t}\, \one_{Y_{t-1}^j \in \Gamma_t}$.
      \item \emph{Resample}: generate a sample
        $( \Yt{t}{j} )_{1 \le j \le m}$ from the distribution
        $\sum_{j=1}^m w_{t,0}^{j} \delta_{\Y{t-1}{j}}$.
      \item \emph{Move}: for each $j \le m$, draw
        $\Y{t}{j} \backsim K\bigl( \Yt{t}{j}, \dotvar \bigr)$. %
        (NB: here, $f$ is evaluated.)
      \end{itemize}
       
    \item Increment~$t$.

    \end{enumerate}

  \item {\bf Estimation} -- Let $m_u$ be the number of particles such that
    $f \bigl( Y_{T-1}^j \bigr) > u$.  Set
    \begin{equation*}
      \alphaSS = \frac{m_u}{m}\, p_0^{T-1}.
    \end{equation*}

  \end{enumerate}
  
  \noindent\rule{\textwidth}{0.4pt}

\end{table}

\endgroup

\section{Bayesian subset simulation}
\label{sec:bss}

\subsection{Bayesian estimation and sequential design of experiment}
\label{sec:seq-design}

Our objective is to build an estimator of~$\alpha$ from the evaluations results
of~$f$ at some points $X_1, X_2, \ldots, X_N \in\XX$, where $N$ is the total
budget of evaluations available for the estimation. In order to design an
efficient estimation procedure, by which we mean both the design of experiments
and the estimator itself, we adopt a Bayesian approach: from now on, the unknown
function~$f$ is seen as a sample path of a random process $\xi$. In other words,
the distribution of~$\xi$ is a \emph{prior} about~$f$. As in
\cite{vaz:09:sysid,bect:2012:stco,chevalier}, the rationale for adopting a
Bayesian viewpoint is to design a good estimation procedure in an average sense.
This point of view has been largely explored in the literature of computer
experiments (see, e.g., \cite{santner:2003:dace}), and that of Bayesian
optimization (see \cite{feliot16:_bayes} and references therein).

For the sake of tractability, we assume as usual that, under the prior
probability that we denote by~$\P_0$, $\xi$ is a Gaussian process (possibly with
a linearly parameterized mean, whose parameters are then endowed with a uniform
improper prior; see \cite{bect:2012:stco} Section~2.3, for details).

Denote by $\E_n$ (resp. $\P_n$) the conditional expectation (resp. conditional
probability) with respect to $X_1, \xi(X_1), \ldots, X_n, \xi(X_n)$, for any
$n \le N$ and assume, as in Section~\ref{sec:subsim}, that $\PX$ has a
probability density function~$\ddpX$ with respect to Lebesgue's measure.  Then,
a natural (mean-square optimal) Bayesian estimator
of~$\alpha = \PX\left( \Gamma \right)$ using $n$ evaluations is the posterior
mean
\begin{equation}
  \label{eq:estimator1}
  \E_n\left( \alpha \right)
  \;=\; \E_n \left( \int_{\XX} \one_{\xi(x)  > u}\, \ddpX(x)\, \dx \right) 
  \;=\; \int_\XX \g_{n,u}(x)\, \ddpX(x)\, \dx,
\end{equation}
where $\g_{n,u}(x) \eqdef \E_n\bigl( \one_{\xi(x) > u} \bigr) = \P_n \bigl(\,
\xi(x) > u \,\bigr)$ is the coverage function of the random set~$\Gamma$ (see,
e.g., \cite{chevalier:moda10}). Note that, since~$\xi$ is Gaussian,
$\g_{n,u}(x)$ can be readily computed for any~$x$ using the kriging equations~(see,
e.g., \cite{bect:2012:stco}, Section~2.4).

Observe that $\g_{n, u} \approx \one_\Gamma$ when the available evaluation
results are informative enough to classify most input points correctly (with
high probability) with respect to~$u$.  This suggests that the computation of
the right-hand side of~\eqref{eq:estimator1} should not be carried out using a
brute force Monte Carlo approximation, and would benefit from an SMC approach
similar to the subset simulation algorithm described in
Section~\ref{sec:subsim}. Moreover, combining an SMC approach with the Bayesian
viewpoint is also beneficial for the problem of choosing (sequentially) the
sampling points~$X_1$, \ldots, $X_N$. In our work, we focus on a \emph{stepwise
  uncertainty reduction} (SUR)
strategy~\cite{vaz:09:sysid,bect:2012:stco}. Consider the function
$L:\widehat \Gamma \mapsto \PX ( \Gamma \symdiff \widehat\Gamma)$, which
quantifies the loss incurred by choosing an estimator $\widehat\Gamma$ instead
of the excursion set $\Gamma$, where $\symdiff$ stands for the symmetric
difference operator. Here, at each iteration $n$, we choose the estimator
$\widehat\Gamma_{n, u} = \left\{ x \in \XX \bmid \g_{n, u} (x) > 1/2
\right\}$.
A SUR strategy, for the loss $L$ and the estimators $\widehat\Gamma_{n, u}$,
consists in choosing a point $X_{n+1}$ at step $n$ in such a way to minimize the
expected loss at step $n+1$:
\begin{equation}
  X_{n+1} = \argmin_{x_{n+1} \in \XX} J_n\left( x_{n+1} \right)\,,
  \label{equ:SUR:argmin}
\end{equation}
where
\begin{equation}
  \label{equ:SUR:Jn}
  J_n\left( x_{n+1} \right) \eqdef \EE_n\bigl(\PX ( \Gamma \symdiff \widehat\Gamma_{n+1, u} )
    \bigm| X_{n+1} = x_{n+1} \bigr)\,.
\end{equation}
For computational purposes, $J_n$ can be rewritten as an integral over $\XX$ of the expected
probability of misclassification $\tau_{n+1,u}$ (see \cite{bect:2012:stco} for more details):
\begin{equation}
  J_n ( x_{n+1} ) =  \int_\XX \EE_n\left( \tau_{n+1,u}(x)\, 
    \bmid X_{n+1} = x_{n+1} \right)\, \ddpX(x)\, \dx.  \label{SUR:integrale}
\end{equation}
where
\begin{equation}
  \label{equ:def:proba-misclass}
  \tau_{n,u} (x) \eqdef %
  \P_n \Bigl( x \in \Gamma \symdiff \widehat\Gamma_{n, u} \Bigr)
    =  \min \Bigl( \g_{n, u}(x),\, 1 - \g_{n, u}(x) \Bigr).
\end{equation}
For moderately small values of~$\alpha$, it is possible to use a sample
from~$\PX$ both for the approximation of the integral in the right-hand side
of~\eqref{SUR:integrale} and for an approximate minimization of~$J_n$ (by
exhaustive search in the set of sample points). However, this simple Monte Carlo
approach would require a very large sample size to be applicable for very small
values of~$\alpha$; a subset-simulation-like SMC approach will now be proposed
as a replacement.

\subsection{A sequential Monte Carlo approach}
\label{sec:optim_strategies}

Assume that~$\alpha$ is small and consider a decreasing sequence of subsets
$\XX = \Gamma_0 \supset \Gamma_1 \supset \cdots \supset \Gamma_T = \Gamma$,
where $\Gamma_t = \left\{ x \in \XX : f(x) > u_t \right\}$, as in
Section~\ref{sec:subsim}. For each~$t \le T$, denote by~$\alphaB_t$ the Bayesian
estimator of~$\alpha_t = \PX\left( \Gamma_t \right)$ obtained from $n_t$
observations of $\xi$ at points $X_1,\,\ldots,\, X_{n_t}$:
\begin{equation}
  \label{eq:estimator2}
  \alphaB_t \;\eqdef\; \E_{n_t}\left( \alpha_t \right)
  \;=\; \int_\XX g_t\, \dPX,
\end{equation}
where $g_t(x) \eqdef \g_{n_t, u_t}(x) = \P_{n_t} \bigl(\, \xi(x) > \u{t} \,\bigr)$.

The main idea of our new algorithm is to use an SMC approach to construct a
sequence of approximations~$\alphaBSS_{t}$ of the Bayesian
estimators~$\alphaB_t$, $1 \le t \le T$ (as explained earlier, the particles of
these SMC approximations will also provide suitable candidate points for the
optimization of a sequential design criterion). To this end, consider the
sequence of probability density functions $q_t$ defined by
\begin{equation}
  \label{equ:qt}
  q_{t}(x) 
  \;\eqdef\; \frac{\ddpX(x)\, g_t(x)}{\int \ddpX(y)\, g_t(y)\, \dy} 
  \;=\; \frac{1}{\alphaB_t}\, \ddpX(x)\, g_t(x).
\end{equation}
We can write a recurrence equation for the sequence of Bayesian
estimators~$\alphaB_t$, similar to that used for the probabilities~$\alpha_t$
in~\eqref{eq:recurr-SS}:
\begin{equation}
  \label{equ:recurr-bss}
  \alphaB_{t+1}
  \;=\; \int g_{t+1}(x)\, \ddpX(x)\, \dx
  \;=\; \alphaB_t\; \int \frac{g_{t+1}(x)}{g_t(x)}\, q_t(x)\, \dx.
\end{equation}
This suggests to construct recursively a sequence of
estimators~$\bigl( \alphaBSS_t \bigr)$ using the following relation:
\begin{equation}
\label{eq:discre}
  \alphaBSS_{t+1} = \alphaBSS_t\;
  \sum_{j=1}^{m} w_t^j\,
  \frac{g_{t+1}(\Y{t}{j})}{g_t(\Y{t}{j})},
  \quad 0 \le t < T,
\end{equation}
where $\bigl( w_t^j, Y_t^j \bigr)_{1 \le j \le m}$ is a weighted sample of
size~$m$ targeting~$q_t$ (as in Section~\ref{sec:sequ-monte-carlo})
and~$\alphaBSS_0 = 1$. The final estimator can be written as:
\begin{equation}
  \label{eq:bss-estimator}
  \alphaBSS_{T}
  \;=\;
  \prod_{t=0}^{T-1} \frac{\alphaBSS_{t+1}}{\alphaBSS_t}
  \;=\;
  \prod_{t=0}^{T-1} \sum_{j=1}^{m} w_t^j\,
  \frac{g_{t+1}(\Y{t}{j})}{g_{t}(\Y{t}{j})}\,.
\end{equation}

\begin{remark}
  The connection between the proposed algorithm and the original subset
  simulation algorithm is clear from the similarity between the recurrence
  relations~\eqref{eq:recurr-SS} and~\eqref{equ:recurr-bss}, %
  and from the use of SMC simulation in both algorithms to construct recursively
  a product-type estimator of the probability of failure (see also this type of
  estimator is mentioned in a very general SMC framework).

  Our choice for the sequence of densities $q_1, \ldots, q_T$ also relates to
  the original subset simulation algorithm. %
  Indeed, note that
  $q_t(x) \propto \E_{n_t} \bigl( \one_{\xi > \u{t}}\, \ddpX \bigr)$, and recall
  from Equation~\eqref{equ:truncated-density} that
  $q_t \propto \one_{\xi > \u{t}}\, \ddpX$ is the target distribution used in
  the subset simulation algorithm at stage~$t$. %
  This choice of instrumental density is also used by
  \cite{dubourg:mbis,dubourg:icasp11} in the context of a two-stage adaptive
  importance sampling algorithm. %
  This is indeed a quite natural choice, since
  $\tilde q_t \propto \one_{\xi > u_t}\, \ddpX$ is the optimal instrumental
  density for the estimation of $\alpha_t$ by importance sampling \citep[see,
  e.g.,][Theorem~3.12]{robert:2004:monte}.%
\end{remark}

\subsection{The Bayesian subset simulation (BSS) algorithm}
\label{sec:implemen}

The algorithm consists of a sequence of stages (or iterations). For the sake of
clarity, assume first that the sequence of thresholds $(u_t)$ is given. Then,
each stage $t\in\NN$ of the algorithm is associated to a threshold $u_t$ and the
corresponding excursion set~$\Gamma_t = \left\{ f > u_t \right\}$.

The initialization stage ($t = 0$) starts with the construction of a \emph{space
  filling} set of points~$\left\{ X_1,\, \ldots,\, X_{n_0} \right\}$ in~$\XX$%
\footnote{%
  See Section~\ref{sec:bss-settings} for more information on the specific
  technique used in this article. %
  Note that it is of course possible, albeit not required to use the BSS
  algorithm, to perform first a change of variables in order to work, e.g., in
  the standard Gaussian space. %
  Whether this will improve the performance of the BSS algorithm is very
  difficult to say in general, and will depend on the example at hand.  }%
, and an initial Monte Carlo sample $\YY{0} = \{ \Y{0}{1}, \ldots, \Y{0}{m} \}$,
consisting of a set of independent random variables drawn from the
density~$q_0 = \ddpX$.

After initialization, each subsequent stage $t \ge 1$ of BSS involves two
phases: an \emph{estimation phase}, where the estimation of $\Gamma_t$ is
carried out, and a \emph{sampling phase}, where a sample~$\YY{t}$ targeting the
density~$q_t$ associated to~$u_t$ is produced from the previous
sample~$\YY{t-1}$ using the reweight/resample/move SMC scheme described in
Section~\ref{sec:reweight--resample}.

In more details, the estimation phase consists in making $N_t \ge 0$ new
evaluations of~$f$ to refine the estimation of $\Gamma_t$. The number of
evaluations is meant to be much smaller than the size~$m$ of the Monte Carlo
sample---which would be the number of evaluations in the classical subset
simulation algorithm.  The total number of evaluations at the end of the
estimation phase at stage~$t$ is denoted by $n_t = n_{t - 1} + N_t$.  The total
number of evaluations used by BSS is thus $n_T = n_0 +\sum_{t = 1}^T N_t$. New
evaluation points $X_{n_{t-1}+1},X_{n_{t-1}+2},\ldots, X_{n_t}$ are determined
using a SUR sampling strategy\footnote{%
  Other sampling strategies (also known as ``sequential design'', or ``active
  learning'' methods) could be used as well. %
  See \citep{bect:2012:stco} for a review and comparison of sampling criterions.
} targeting the threshold~$\u{t}$, as in Section~\ref{sec:seq-design} (see
Supplementary Material~\ref{SM:SUR-crit} for details about the numerical
procedure).

In practice, the sequence of thresholds is not fixed beforehand and adaptive
techniques are used to choose the thresholds (see
Section~\ref{subsec:adapt-thresh}) and the number of points per stage (see
Section~\ref{subsec:autoSUR}).

The BSS algorithm is presented in pseudo-code form in Table~\ref{tab:imple}.

\begin{remark}
  Algorithms involving Gaussian-process-based adaptive sampling and subset
  simulation have been proposed by Dubourg and co-authors
  \cite{Dubourg:2011:phd, dubourg:2011:rbdo} and by Huang et
  al.~\cite{Huang2016}.  Dubourg's work addresses a different problem (namely,
  reliability-based design optimization).  Huang et al.'s paper, published very
  recently, adresses the estimation of small probabilities of failure.  We
  emphasize that, unlike BSS, none of these algorithms involves a direct
  interaction between the selection of evaluation points (adaptive sampling) and
  subset simulation---which is simply applied, in its original form, to the
  posterior mean of the Gaussian process (also known as kriging predictor).
\end{remark}

\begingroup
\linespread{0}

\begin{table}
  \SetTableFontSize

  \caption{Bayesian subset simulation algorithm}
  \label{tab:imple}

  \noindent\rule{\textwidth}{0.4pt}
  \begin{enumerate}[itemsep=2ex]
  \item {\bf Initialization } (stage $0$)
    \begin{enumerate}[topsep=0pt, itemsep=0pt]
    \item Evaluate~$f$ on a set of points $\{X_1$, \ldots, $X_{n_0}\}$, called
      the \emph{initial design} (see Section~\ref{sec:bss-settings} for details)
    \item Generate an \iid sample $\YY{t} = \{ \Y{0}{1}, \ldots, \Y{0}{m}\}$
      from $\PX$.
    \item Choose a prior $\P_0$ (see Sections~\ref{sec:seq-design}
      and~\ref{sec:bss-settings} for details).
    \item Set $u_0 = -\infty$, $g_0 = \g_{0,-\infty} = \one_\XX$, $n = n_0$
      and~$t = 1$.
    \end{enumerate}
    
    \medbreak
    
  \item {\bf Repeat } (stage~$t$)
    \begin{enumerate}
    \item {\bf Estimation}
      \begin{itemize}
      \item Set $k = 0$ and repeat
        \begin{itemize}
        \item Select a threshold $\tilde u_{t,k}$ by solving
          Equation~\eqref{eq:p0-select-rule} for~$u_t$ (with $n_t = n$)
        \item Stop if the condition~\eqref{equ:stopping-crit} is met, with $n_t
          = n$ and $u_t = \tilde u_{t,k}$.
        \item Select~$X_{n+1}$ using the SUR strategy
          \eqref{equ:SUR:argmin}--\eqref{equ:def:proba-misclass} with respect
          to~$\tilde u_{t,k}$.
        \item Evaluate~$f$ at~$X_{n+1}$. Increment~$n$ and~$k$.
        \end{itemize}
      \item Set $N_t = k$, $n_t = n$, $u_t = \tilde u_{t,k}$ and %
        $g_{t} = \g_{n_t, u_t} = \P_{n_t} \bigl(\, \xi\left( \dotvar
        \right)) > \u{t} \,\bigr)$.
      \end{itemize}
    \item {\bf Sampling}
      \begin{flushleft}
        \begin{itemize}
        \item \emph{Reweight}: calculate weights $w_{t,0}^{j} \propto
          g_{t}(\Y{t-1}{j}) / g_{t-1}(\Y{t-1}{j})$, $1 \le j \le m$.
        \item \emph{Resample}: generate a sample $( \Yt{t}{j} )_{1 \le j \le m}$ from the
          distribution $\sum_{j=1}^m w_{t,0}^{j} \delta_{\Y{t-1}{j}}$.
          \vspace{-1mm}
        \item \emph{Move}: for each $j \le m$, draw $\Y{t}{j}
          \backsim K\bigl( \Yt{t}{j}, \dotvar \bigr)$.
        \end{itemize}
      \end{flushleft}
    \item Increment~$t$.
    \end{enumerate}

  \item {\bf Estimation} -- The final probability of failure is estimated by
    \begin{equation*}
      \alphaBSS_T = \prod_{t=0}^{T-1} \left(
        \frac{1}{m} \sum_{j=1}^{m} \frac{g_{t+1}(\Y{t}{j})}{g_{t}(\Y{t}{j})}
      \right).
    \end{equation*}
    
  \end{enumerate}

  \noindent\ignorespaces\rule{\textwidth}{.4pt}%

\end{table}

\endgroup

\subsection{Adaptive choice of the thresholds $u_{t}$}
\label{subsec:adapt-thresh}

As discussed in Section~\ref{sec:subsim-adapt-thresh}, it can be proved that,
for an idealized version of the subset simulation algorithm with fixed
thresholds $\u{0} < \u{1} < \cdots < \u{T} = \u{}$, it is optimal to choose the
thresholds to make all conditional probabilities
$\PX\bigl( \Gamma_{t+1} | \Gamma_{t} \bigr)$ equal.  This leads to the idea of
choosing the thresholds adaptively in such a way that, in the product estimate
\begin{equation*}
  \alphaSS_T \;=\;
  \prod_{t=1}^T \frac{1}{m} \sum_{i=1}^m
  \one_{\Gamma_t}\bigl( \Y{t-1}{i} \bigr),
\end{equation*}
each term but the last is equal to some prescribed constant~$p_0$. In other
words, $\u{t}$ is chosen as the $(1-p_0)$-quantile of~$\YY{t-1}$. This idea was
first suggested by (\cite{au01:_estim}, Section~5.2), on the heuristic ground
that the algorithm should perform well when the conditional probabilities are
neither too small (otherwise they are hard to estimate) nor too large (otherwise
a large number of stages is required).

Consider now an idealized BSS algorithm, where a) the initial design of
experiment is independent of~$\YY{0}$, b) the SUR criterion is computed exactly,
or using a discretization scheme that does not use the~$\YY{t}$'s; c) the
minimization of the SUR criterion is carried out independently of the~$\YY{t}$'s
and d) the particles $Y_t^j$ are independent and identically distributed
according to~$q_t$. Assumptions~a)--c) ensure that the sequence of densities
$(q_t)_{1 \le t \le T}$ is deterministic given~$\xi$. Then (see
Appendix~\ref{app:variance}),
\begin{equation}
  \label{eq:var-bss-1}
  \var \left(
    \frac{\alphaBSS_T}{\alphaB_T}
    \Biggm| \xi
  \right)
  =
  \frac{1}{m}\,
  \sum_{t = 1}^T \kappa_t
  + O\left( \frac{1}{m^2} \right),
\end{equation}
where
\begin{equation}
  \label{eq:var-bss-2}
  \kappa_t \eqdef
  \frac{\int_\XX g_t^2 / g_{t-1}\, \ddpX}{%
    \left( \alphaB_t \right)^2 / \alphaB_{t - 1}}
  - 1.
\end{equation}
Minimizing the leading term $\frac{1}{m} \sum_{t = 1}^T \kappa_t$
in~\eqref{eq:var-bss-1} by an appropriate choice of thresholds is not as
straightforward as in the case of the subset simulation algorithm. Assuming that
$g_{t-1} \approx 1$ wherever $g_t$ is not negligible (which is a reasonable
assumption, since $g_t(x) = \P_{n_t} \bigl(\, \xi(x) > \u{t} \,\bigr)$ and
$\u{t} > \u{t-1}$), we get
\begin{equation*}
  \int_\XX g_t^2 / g_{t-1}\, \ddpX
  \approx
  \int_\XX g_t^2\, \ddpX
  \le 
  \int_\XX g_t\, \ddpX
  =
  \alphaB_t,
\end{equation*}
and therefore the variance~\eqref{eq:var-bss-1} is approximately upper-bounded by
\begin{equation}
  \label{eq:upper-bound-var}
  \frac{1}{m} \sum_{t= 1}^T \left( 1 - \pB_t \right) / \pB_t,
\end{equation}
where $\pB_t \eqdef \alphaB_t / \alphaB_{t - 1}$. Minimizing the approximate
upper-bound~\eqref{eq:upper-bound-var} under the constraint
\begin{equation*}
  \prod_{t=1}^T \pB_t = \alphaB_T  
\end{equation*}
leads to choosing the thresholds~$u_t$ in such a way that $\pB_t$ is the same
for all stages~$t$---that is $\pB_t = \left(\alphaB_T\right)^{1/T}$. As a
consequence, we propose to choose the thresholds adaptively using the condition
that, at each stage (but the last), the natural estimator
$\alphaBSS_t / \alphaBSS_{t - 1}$ of~$\pB_t$ is equal to some prescribed
probability~$p_0$. Owing to~(\ref{eq:discre}), this amounts to choosing $u_t$ in
such a way that
\begin{equation}
  \label{eq:p0-select-rule}
  \frac{1}{m} \sum_{i=1}^{m} \frac{%
    g_t(\Y{t - 1}{i})%
  }{%
    g_{t - 1}(\Y{t - 1}{i})%
  } 
  = p_0.
\end{equation}
should be satisfied.

Equation~\eqref{eq:p0-select-rule} is easy to solve, since the left-hand side is
a strictly decreasing and continuous function of~$\u{t}$ (to be precise,
continuity holds under the assumption that the posterior variance of~$\xi$ does
not vanish on one of the particles).  In practice, we
solve~\eqref{eq:p0-select-rule} each time a new evaluation is made, which yields
a sequence of intermediate thresholds (denoted by~$\tilde u_{t,0}$,
$\tilde u_{t,1}$\ldots\ in Table~\ref{tab:imple}) at each stage~$t\ge 1$.  The
actual value of $u_t$ at stage~$t$ is only known after the last evaluation of
stage~$t$.

\begin{remark}
  Alternatively, the effective sample size (ESS) could be used to
  select the thresholds, as proposed by~\cite{DelMoral2012}.  This
  idea will not be pursued in this paper.  The threshold selected by
  the ESS-based approach will be close to the threshold selected by
  Equation~\eqref{eq:p0-select-rule} when the all the ratios~$g_t(\Y{t
    - 1}{i}) / g_{t - 1}(\Y{t - 1}{i})$, or most of them, are either
  close to zero or close to one.
\end{remark}

\subsection{Adaptive choice of the number~$N_t$ of evaluation at each stage}
\label{subsec:autoSUR}

In this section, we propose a technique to choose adaptively the number~$N_t$ of
evaluations of~$f$ that must be done at each stage of the algorithm.

Assume that~$t \ge 1$ is the current stage number; at the beginning of the
stage, $n_{t-1}$ evaluations are available from previous stages. After several
additional evaluations, the number of available observations of~$f$ is
$n \ge n_{t-1}$.  We propose to stop adding new evaluations when the expected
error of estimation of the set~$\Gamma_t$, measured by
$\EE_n \left( \P_\XX \left( \Gamma_t \symdiff \widehat\Gamma_{n,u_t} \right)
\right)$,
becomes smaller than some prescribed fraction~$\eta_t$ of its expected volume
$\EE_n \left( \P_\XX \left( \Gamma_t \right) \right)$ under~$\P_\XX$. Writing
these two quantities as
\begin{align*}
  \EE_n \left( \P_\XX \left( \Gamma_t \right) \right) 
  & =\;
  \int_\XX \g_{n, u_t} (x)\, \ddpX(x)\, \dx
  = \alphaB_{t-1}\, \int_\XX \frac{\g_{n, u_t} (x)}{g_{t-1}(x)} q_{t-1}(x)\, \dx,
  \\
  \EE_n \left( \P_\XX \left(  \Gamma \symdiff \widehat\Gamma_{n, u_t} \right) \right) 
  & =\;
  \int_\XX \tau_{n,u_t} (x)\, \ddpX(x)\, \dx
  = \alphaB_{t-1}\, \int_\XX \frac{\tau_{n,u_t} (x)}{g_{t-1}(x)} q_{t-1}(x)\, \dx,
\end{align*}
where~$\g_{n, u_t}$ and~$\tau_{n,u_t}$ have been defined in
Section~\ref{sec:seq-design}, and estimating the integrals on the right-hand
side using the SMC sample~$\YY{t-1}$, we end up with the stopping condition \def
\tmp {\left( \Y{t-1}{i} \right)}
\begin{equation*}
  \frac{1}{m}\, \sum_{i=1}^m \frac{\tau_{n,u_t} \tmp}{g_{t-1} \tmp}
  \;\le\; \eta_t\,
  \cdot
  \frac{1}{m}\, \sum_{i=1}^m \frac{\g_{n, u_t} \tmp}{g_{t-1} \tmp}.
\end{equation*}
which, if $u_t$ is re-adjusted after each evaluation using
Equation~\eqref{eq:p0-select-rule}, boils down to
\begin{equation}
  \label{equ:stopping-crit}
  \sum_{i=1}^m \frac{\tau_{n,u_t} \tmp}{g_{t-1} \tmp}
  \;\le\; \eta_t m p_0.
\end{equation}

\begin{remark}
  In the case where several evaluations of the function can be carried out in
  parallel, it is possible to select evaluation points in batches during the
  sequential design phase of the algorithm. A batch-sequential version of the
  SUR strategy~\eqref{equ:SUR:argmin}--\eqref{equ:def:proba-misclass} has been
  proposed by \cite{chevalier}.
\end{remark}

\begin{remark}
  The stopping criterion~\eqref{equ:stopping-crit} is slightly different from
  the one proposed earlier by \cite{lilyPSAM}:
  $\sum_{i=1}^m \tau_{n,u_t} \tmp \le \eta' m$. If we set $\eta' = \eta_t p_0$
  and assume (quite reasonably) that $g_{t - 1} \tmp \approx 1$ for the
  particles where $\tau_{n,u_t} \tmp$ is not negligible, then it becomes clear
  that the two criterions are essentially equivalent. As a consequence, the
  left-hand side of~\eqref{equ:stopping-crit} can also be interpreted,
  approximately, as the expected number of misclassified particles (where the
  expectation is taken with respect to~$\xi$, conditionally on the particles).
\end{remark}

\section{Numerical experiments}
\label{sec:num}

\def \alphaRefA {$5.596\!\cdot\! 10^{-9}$}
\def \alphaRefB {$3.937\!\cdot\! 10^{-6}$}
\def \alphaRefC {$1.514\!\cdot\! 10^{-8}$}

In this section, we illustrate the proposed algorithm on three classical
examples from the structural reliability literature and compare our results with
those from the classical subset simulation algorithm and the \TwoSMART algorithm
\cite{deheeger:2008, bourinet2011assessing}. %
These examples are not actually expensive to evaluate, which makes it possible
to analyse the performance of the algorithms through extensive Monte Carlo
simulations, but the results are nonetheless relevant to case of
expensive-to-evaluate simulators since performance is measured in terms of
number of function evaluations (see Section~\ref{sec:results:average} for a
discussion).

The computer programs used to conduct these numerical experiments are freely
available from \url{https://sourceforge.net/p/kriging/contrib-bss} under the LGPL
licence~\cite{lgplv21}. %
They are written in the Matlab/Octave language and use the STK
toolbox~\cite{stktoolbox} for Gaussian process modeling. %
For convenience, a software package containing both the code for the BSS
algorithm itself and the STK toolbox is provided as Supplementary Material.

\begin{table}
  \SetTableFontSize\centering

  \caption{Summary of test cases}
  \label{tab:summary-test-cases}

  \def\vv{\vphantom{$\Big|$}}
  \begin{tabular}{|c|l|c|c|}
    \hline
    Example & Name & $d$ & $\alpha_{\text{ref}}$\\
    \hline
    \ref{ex:FB}  & Four-branch series system          & $2$ & \alphaRefA\vv\\
    \ref{ex:DCB} & Deviation of a cantilever beam     & $2$ & \alphaRefB\\
    \ref{ex:RNO} & Response of a nonlinear oscillator & $6$ & \alphaRefC\vv\\
    \hline
  \end{tabular}

\end{table}

\subsection{Test cases}
\label{sec:test-cases}

For each of the following test cases, the reference value for the
probability~$\alpha$ has been obtained from one hundred independent runs of the
subset simulation algorithm with sample size $m = 10^7$ (see
Table~\ref{tab:summary-test-cases}).

\subsubsection{Four-branch series system} \label{ex:FB} 
\begin{figure}[ht!]
  \centering
  \includegraphics{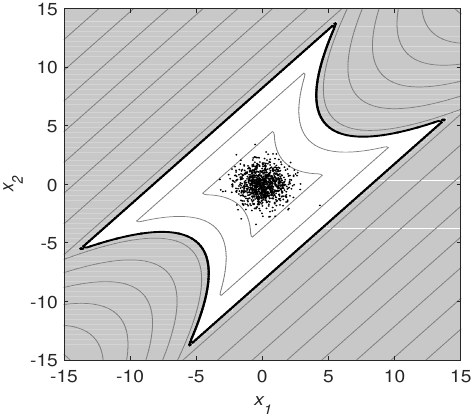}
  \hspace{1cm}
  \includegraphics{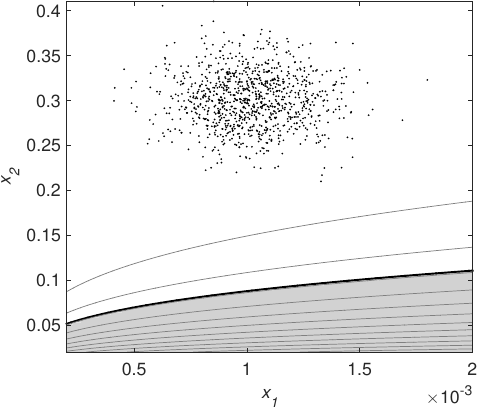}
  \caption{Contour plots of $f$ in Example~\ref{ex:FB} (left) and
    Example~\ref{ex:DCB} (right), along with a sample of size~$m=10^{3}$
    from~$\PX$~(dots). A failure happens when~$x$ is in the gray
    area.}  \label{fig:contour-plots}
\end{figure}

Our first example is a
variation on a classical structural reliability test case (see,
e.g., \cite{echard:2011:akmcs}, Example~1, with~$k=6$), where the threshold~$u$ is
modified to make~$\alpha$ smaller. The objective is to estimate the probability
$\alpha = \PX \left( f(X) < u \right)$, where
\begin{equation}
  f(x_1,x_2) = \min \left\{
    \begin{array}{ll}
      3+0.1(x_1-x_2)^{2}-(x_1+x_2)/ \sqrt{2}, \\
      3+0.1(x_1-x_2)^{2}+(x_1+x_2)/ \sqrt{2}, \\
      (x_1-x_2)+6/ \sqrt{2}, \\
      (x_2-x_1)+6/ \sqrt{2}
    \end{array} \right \}
\end{equation}
and~$X_1, X_2 \stackrel{\text{iid}}{\sim} \mathcal{N}(0,1)$. Taking $\u{} = -4$,
the probability of failure is approximately~\alphaRefA, with a coefficient of
variation of about $0.04\%$. Figure~\ref{fig:contour-plots} (left panel) shows
the failure domain and a sample from the input distribution~$\PX$.

\subsubsection{Deviation of a cantilever beam} \label{ex:DCB} Consider a
cantilever beam, with a rectangular cross-section, subjected to a uniform
load. The deflection of the tip of the beam can written as
\begin{equation}
  f(x_{1}, x_{2}) = \frac{3L^4}{2E}\frac{x_1}{x_2^3}\,,
\end{equation}
where $x_1$ is the load per unit area, $x_2$ the thickness of the beam,
$L=6\,{\rm m}$ and $E=2.6\!\cdot\!10^{4}\,{\rm MPa}$. The input variable~$X_1$
and~$X_2$ are assumed independent, with
$X_1 \sim \Ncal\bigl(\mu_1, \sigma_1^2)$, $\mu_1=10^{-3}\,{\rm MPa}$,
$\sigma_1=0.2\mu_1$, and $X_{2} \sim \Ncal\bigl(\mu_2,\sigma_2^2\bigr)$,
$\mu_2=0.3\,{\rm m}$, $\sigma_2=0.1\mu_2$. A failure occurs when $f$ is larger
than $\u{} = L/325$. The probability of failure is approximately~\alphaRefB,
with a coefficient of variation of about $0.03\%$. Note that the distribution
of~$X_2$ has been modified, with respect to the usual formulation (see, e.g.,
\cite{gayton2003cq}), to make~$\alpha$ smaller.  Figure~\ref{fig:contour-plots}
(right panel) shows a contour plot of $f$, along with a sample of the input
distribution.

\subsubsection{Response of a nonlinear oscillator} \label{ex:RNO}

In this example (see, e.g., \cite{Echard:2013}), the input variable is
six-dimensional and the cost function is:
\begin{equation}
  f \left( x_1, x_2, x_3, x_4, x_5, x_6 \right) =
  3 x_{4} - \left\lvert \frac{2x_{5}}{x_{1}{w_{0}}^{2}}
    \sin\Big( \frac{w_{0}x_{6}}{2}    \Big) \right\rvert
\end{equation}
where $w_{0} = \sqrt{\frac{x_{2}+x_{3}}{x_{1}}}$. The input variables are
assumed independent, normal, with mean and variance parameters given in
Table~\ref{tab:ex3:var}. A failure happens when the cost function is lower than
the threshold~$\thres = 0$. The probability of failure is
approximately~\alphaRefC, with a coefficient of variation of about
$0.04\%$. This variant of the problem corresponds exactly to the harder case
in~\cite{Echard:2013}.

\begin{table}  
  \SetTableFontSize\centering

  \caption{Example~\ref{ex:RNO}: Means and standard deviations of the input variables.}
  \label{tab:ex3:var}

  \begin{tabular}{|c|c c c c c c|}
    \hline
    Variable & $x_{1}$ & $x_{2}$ & $x_{3}$ & $x_{4}$ & $x_{5}$ & $x_{6}$\\
    \hline
    $\mu_i$ & $1$ & $1$ & $0.1$ & $0.5$ & $0.45$ & $1$\\
    $\sigma_i$ & $0.05$ & $0.1$ & $0.01$ & $0.05$ & $0.075$ & $0.2$\\
    \hline
  \end{tabular}

\end{table}

\subsection{Experimental settings}
\label{sec:settings}

\subsubsection{BSS algorithm} \label{sec:bss-settings}

\paragraph{Initial design of experiments} We start with an initial design of
size~$n_0 = 5 d$ (see \cite{loeppky2009css} for a discussion on the size of the
initial design in computer experiments), generated as follows. First, a compact
subset~$\XX_0 \subset \XX$ is constructed\footnote{A similar technique is used
  by Dubourg and co-authors in a context of reliability-based design
  optimization \cite{dubourg:2011:rbdo, Dubourg:2011:phd}.}:
\begin{equation*}
  \XX_0 = \prod_{j=1}^d \left[ q_\varepsilon^j; q_{1 - \varepsilon}^j \right]
\end{equation*}
where $q_\varepsilon^j$ and~$q_{1 - \varepsilon}^j$ are the quantiles of
order~$\varepsilon$ and~$1 - \varepsilon$ of the~$j^{\text{th}}$ input variable.
Then, a ``good'' LHS design on~$\left[ 0; 1 \right]^d$ is obtained as the best
design according the maximin criterion \cite{JMY90, MM95} in a set of $Q$ random
LHS designs, and then scaled to~$\XX_0$ using an affine mapping. The
values~$\varepsilon = 10^{-5}$ and~$Q = 10^4$ have been used in all our
experiments.

\paragraph{Stochastic process prior} A Gaussian process prior with an unknown
constant mean and a stationary anisotropic Mat\'ern covariance function with
regularity~$5/2$ is used as our prior information about~$f$ (see Supplementary
Material~\ref{SM:spp} for more details). The unknown mean is integrated out as
usual, using an improper uniform prior on~$\RR$; as a consequence, the posterior
mean coincides with the so-called ``ordinary kriging'' predictor. The remaining
hyper-parameters (variance and range parameters of the covariance function) are
estimated, following the empirical Bayes philosophy, by maximization of the
marginal likelihood\footnote{Used in combination with a uniform prior for the
  mean, for this specific model, the MML method is equivalent to the Restricted
  Maximum Likelihood (ReML) method advocated by \cite{Ste99}, Section~6.4.}. The
hyper-parameters are estimated first on the data from the initial design, and
then re-estimated after each new evaluation. %
In practice, we recommend to check the estimated parameters every once in a
while using, e.g., leave-one-out cross-validation.

\paragraph{SMC parameters} Several values of the sample size~$m$ will be tested
to study the relation between the variance of the estimator and the number of
evaluations: $m \in \{ 500, 1000, 2000, \ldots \}$. Several iterations of an
adaptive Gaussian Random Walk Metropolis-Hastings (RWMH) algorithm, fully
described in Supplementary Material~\ref{SM:adaptiveMCMC}, are used for the move
step of the algorithm.

\paragraph{Stopping criterion for the SUR strategy} The number of evaluations
selected using the SUR strategy is determined adaptively, using the stopping
criterion~\eqref{equ:stopping-crit} from Section~\ref{subsec:autoSUR}, with
$\eta_t = 0.5$ for all~$t < T$ (i.e., for all intermediate stages)
and~$\eta_T = 0.1\, \hat\delta_{m, T}$ where~$\hat\delta_{m, T}$ is the
estimated coefficient of variation for the SMC estimator~$\alphaBSS_T$
of~$\alphaB_T$ (see Appendix~\ref{app:variance}). Furthermore, we require for
robustness a minimal number~$N_{\text{min}}$ of evaluations at each stage,
with~$N_{\text{min}} = 2$ in all our simulations.

\paragraph{Adaptive choice of the thresholds} The successive thresholds~$u_t$
are chosen using the adaptive rule proposed in
Section~\ref{subsec:adapt-thresh}, Equation~\eqref{eq:p0-select-rule},
with~$p_{0} = 0.1$. %
This value has been found experimentally to be neither too large (to avoid
having a large number of stages) not too small (to avoid losing too many
particles during the resampling step)\footnote{Note that the considerations of
  Remark~\ref{rem:choix-p0} on the choice of~$p_0$ are not relevant here, since
  the computational cost of our method is mainly determined by the number of
  function evaluations, which is not directly related to the number of particles
  to be simulated.  See also Supplementary Material~\ref{SM:runtime}.}.

\subsubsection{Subset simulation algorithm}

The parameters used for the subset simulation algorithm are exactly the same, in
all our simulations, as those used in the ``SMC part'' of the BSS algorithm (see
Section~\ref{sec:bss-settings}). In particular, the number~$m_0$ of surviving
particles at each stage is determined according to the
rule~$p_0 = \frac{m_0}{m} = 0.1$ (see Table~\ref{tab:subsim_algo}), and the
adaptive MCMC algorithm described in Supplementary
Material~\ref{SM:adaptiveMCMC} is used to move the particles.  The number of
evaluations made by the subset simulation algorithm is considered to be
$m + \left(T-1\right) \left(1-p_0\right) m$, as explained in
Section~\ref{sec:subsim-adapt-thresh}---in other words, in order to make the
comparison as fair as possible, the additional evaluations required by the
adaptive MCMC procedure are not taken into account.

\subsubsection{\TwoSMART algorithm}
\label{sec:settings:2smart}

\TwoSMART \cite{bourinet2011assessing, deheeger:2008} is another algorithm for
the estimation of small probabilities, which is based on the combination of
subset simulation with Support Vector Machines (SVM). We will present results
obtained using the implementation of \TwoSMART proposed in the software package
FERUM~4.1 \cite{bourinet2010}, with all parameters set to their default values
(which are equal to the values given in~\cite{bourinet2011assessing}).

\begin{remark}
  Several other methods addressing the estimation of small probabilities of
  failure for expensive-to-evaluate functions have appeared recently in the
  structural reliability literature~\cite{Balesdent2013, Bourinet2016,
    Cadini2014, Echard:2013, Huang2016}.  \TwoSMART was selected as a reference
  method due to the availability of a free software implementation in FERUM.  A
  more comprehensive benchmark is left for future work.
\end{remark}

\subsection{Results}
\label{sec:results}

\subsubsection{Illustration} \label{sec:results:illustr} %
\begin{figure}
  \centering
  \vspace{1cm}
  \includegraphics{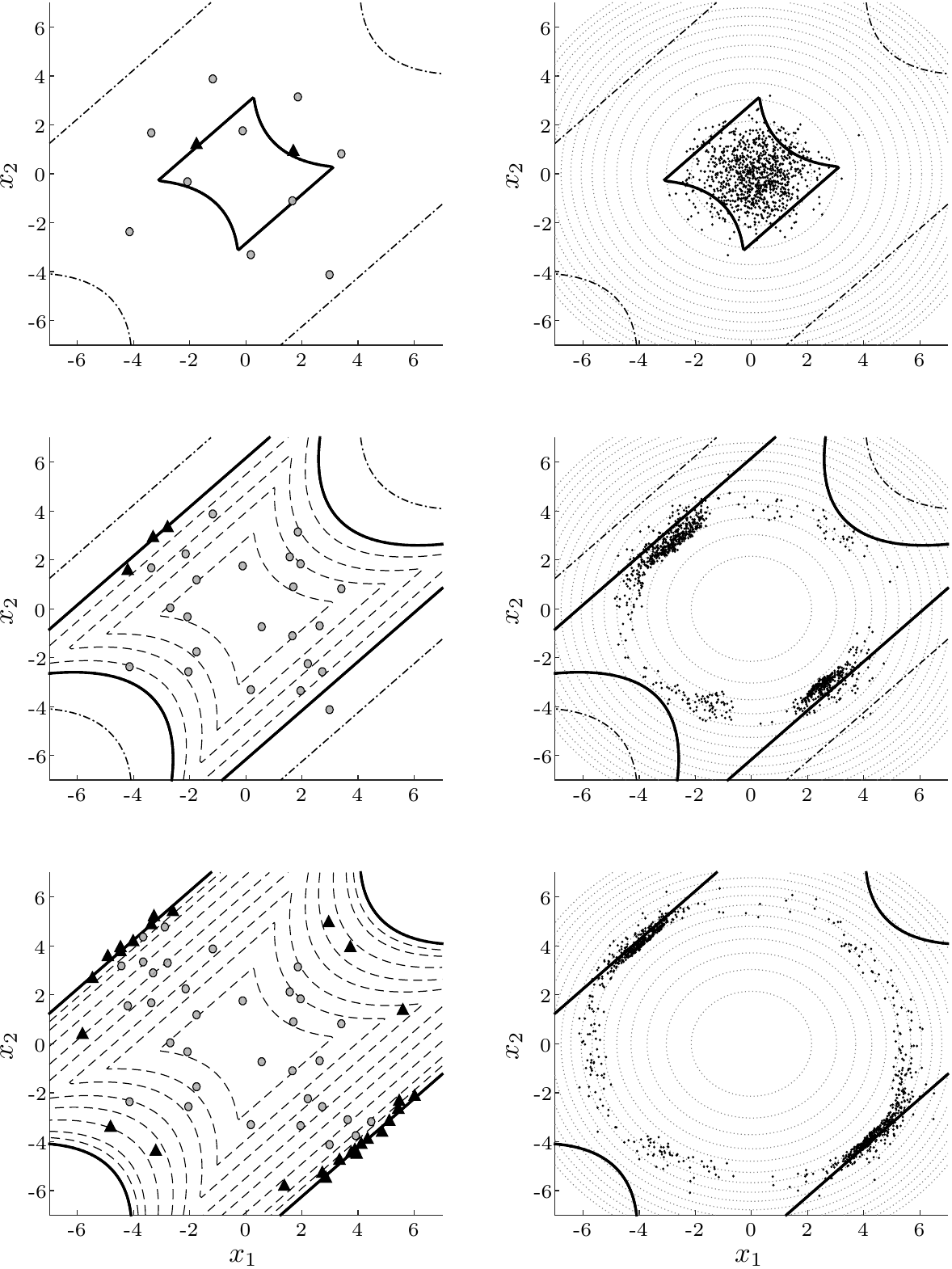}  
  \smallskip
  \caption{Snapshots of the BSS algorithm running on Example~\ref{ex:FB} (four
    branch series system) with sample size~$m = 1000$. The first, second and
    third row correspond respectively to the end of the first stage ($t = 1$),
    the fifth stage ($t = 5$) and the last stage ($t = T = 9$). The true level
    set corresponding current target level~$u_t$ is represented by a thick line
    and, in the left column, true level sets corresponding to previous levels
    ($u_s$, $s < t$) are recalled using dashed contours. Evaluation points from
    previous stages are represented by gray disks (in particular, the initial
    design of experiment of size~$n_0 = 10$ is visible on the top-left panel)
    and new evaluations performed at the current levels are marked by black
    triangles. In the right column, the sample points~$Y_{t-1}^j$,
    $1 \le j \le m$ and the level sets of the input density~$\ddpX$
    (corresponding to probabilities~$1 - 10^{-k}$, $k = 1, 2, \ldots$) are
    represented respectively by black dots and dotted lines.}
  \label{fig:fourbranch-illustr}
\end{figure}

We first illustrate how BSS works using one run of the algorithm on
Example~\ref{ex:FB} with sample size~$m = 1000$. Snapshots of the algorithm at
stages~$t = 1$, $t = 5$ and~$t = T = 9$ are presented on
Figure~\ref{fig:fourbranch-illustr}. Observe that the additional evaluation
points selected at each stage using the SUR criterion (represented by black
triangles) are located in the vicinity of the current level set. The actual
number of points selected at each stage, determined by the adaptive stopping
rule, is reported in Table~\ref{tab:nbEvalsPerStage}. Observe also that the set
of particles (black dots in the right column) is able to effectively capture the
bimodal target distribution. Finally, observe that a significant portion of the
evaluation budget is spent on the final stage---this is again a consequence of
our adaptive stopping rule, which refines the estimation of the final level set
until the bias of the estimate is (on average under the posterior distribution)
small compared to its standard deviation.

\begin{table}[ht!]
  \SetTableFontSize\centering

  \caption{Number of evaluations per stage on Example~\ref{ex:FB} (four-branch series
    system). For the BSS algorithm, recall that the number of evaluations at
    each stage is chosen adaptively (see Section~\ref{subsec:autoSUR}) and is
    therefore random: the numbers shown here correspond to the run with~$m = 1000$
    that is shown on Figure~\ref{fig:fourbranch-illustr}. For the subset
    simulation algorithm, the number of evaluations is directly related to
    the~$m$, $q_0 = 1 - p_0$ and~$T$ (see
    Section~\ref{sec:subsim-adapt-thresh}).}
  \label{tab:nbEvalsPerStage}

  \begin{tabular}{|c|c|c|c|c|c|c|}
    \hline
    stage number~$t$ %
    & $0$ & $1$ & $2$  & $3$ & $4$ & $5$
    \\ \hline
    BSS ($m = 1000$) %
    & $2d = 10$ & $2$ & $6$ & $3$ & $2$ & $3$
    \\ \hline
    subset simulation %
    & $m$ & $q_0 m$ & $q_0 m$ & $q_0 m$ & $q_0 m$ & $q_0 m$
    \\ \hline
  \end{tabular} \smallskip

  \begin{tabular}{|c|c|c|c|c|c|}
    \hline
    stage number~$t$ %
    & $6$ & $7$ & $8$ & $9$ & total
    \\ \hline
    BSS ($m = 1000$) %
    & $2$ & $3$ & $2$ & $28$ & $61$
    \\ \hline
    subset simulation %
    & $q_0 m$ & $q_0 m$ & $q_0 m$ & 0  & $m + (T-1) q_0 m$
    \\ \hline
  \end{tabular}

\end{table}

\subsubsection{Average results} \label{sec:results:average} %
This section presents average results over one hundred independent runs for
subset simulation, BSS and~\TwoSMART.

\begin{figure}
  \centering  

  \subfloat[Example~\ref{ex:FB}: Four branch]{%
    \includegraphics{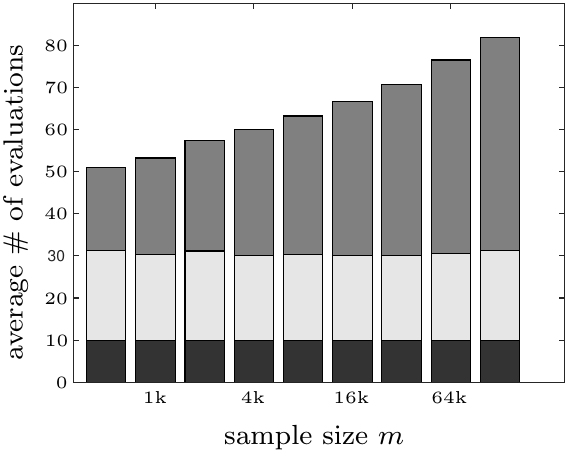}}%
  \hspace{5mm}
  \subfloat[Example~\ref{ex:DCB}: Cantilever beam]{%
    \includegraphics{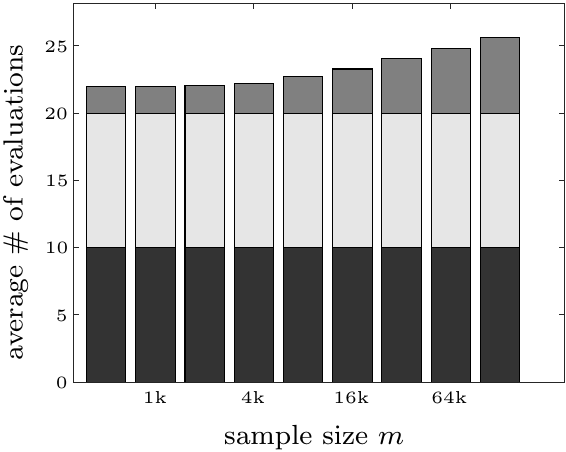}}%

  \smallskip 

  \subfloat[Example~\ref{ex:RNO}: Nonlinear oscillator]{%
    \includegraphics{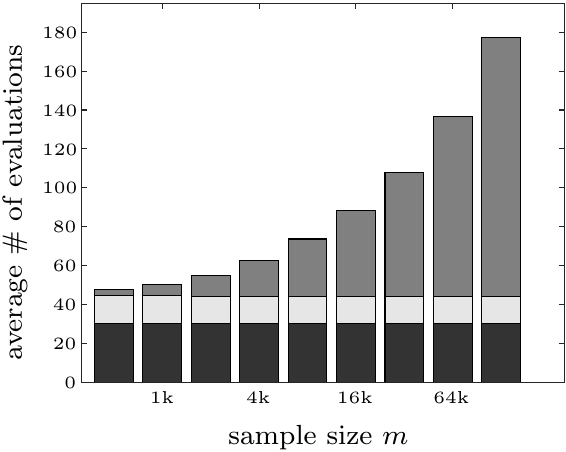}}%

  \smallskip

  \caption{Average number of evaluations used by the BSS algorithm, over 100
    independent runs, as a function of the sample size~$m$ on
    Examples~\ref{ex:FB}--\ref{ex:RNO}.  The total number of evaluations is
    split in three parts: the size~$n_0$ of the initial design (dark gray), the
    number $\sum_{t=1}^{T-1} N_t$ of evaluations in intermediate stages (light
    gray) and the number of evaluations~$N_T$ in the final stage (middle gray).}
  \label{fig:nbEvalsHisto}
\end{figure}

Figure~\ref{fig:nbEvalsHisto} represents the average number of evaluations used
by the BSS algorithm as a function of the sample size~$m$. The number of
evaluations spent on the initial design is constant, since it only depends on
the dimension~$d$ of the input space. The average number of evaluations spent on
the intermediate stages ($t < T$) is also very stable\footnote{Actually, for
  Examples~\ref{ex:DCB} and~\ref{ex:RNO}, it is equal to~$T\, N_{\text{min}}$
  for all runs; in other words, the adaptive stopping rule only came into play
  at intermediate stages for Example~\ref{ex:FB}.} and independent of the sample
size~$m$. Only the average number of evaluations spent on the final
stage---i.e., to learn the level set of interest---is growing with~$m$. This
growth is necessary if one wants the estimation error to decrease when~$m$
increases: indeed, the variance of~$\alphaBSS_T$ automatically goes to zero at
the rate~$\frac{1}{m}$, but the bias~$\alphaB_T - \alpha$ does not unless
additional evaluations are added at the final level to refine the estimation
of~$\Gamma$.

\begin{figure}
  \centering

  \subfloat[Example~\ref{ex:FB}: Four branch]{%
    \includegraphics{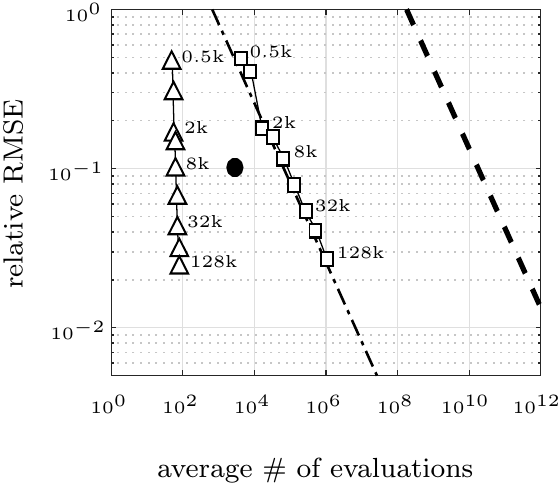}}%
  \hspace{5mm}
  \subfloat[Example~\ref{ex:DCB}: Cantilever beam]{%
    \includegraphics{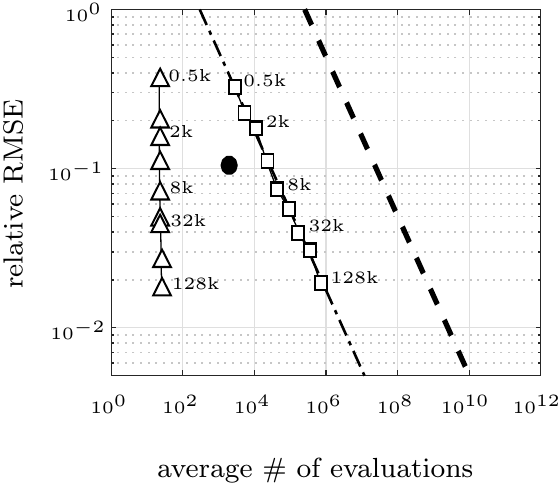}}%

  \smallskip 

  \subfloat[Example~\ref{ex:RNO}: Nonlinear oscillator]{%
    \includegraphics{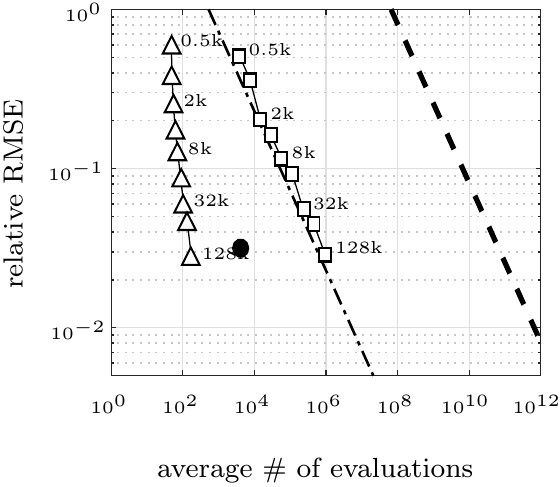}}%

  \smallskip

  \caption{Relative root-mean-square error (RMSE) as a function of the average
    number of evaluations, over 100 independent runs, on
    Examples~\ref{ex:FB}--\ref{ex:RNO}. For the subset simulation algorithm
    (squares) and for the BSS algorithm (triangles), the results are provided
    for several values of the sample size ($m \in \{500, 1000, 2000,
    \ldots\}$). For the \TwoSMART algorithm (filled circles), only one result is
    presented, corresponding to the default settings of the algorithm. The
    expected performance of plain Monte Carlo sampling is represented by a
    dashed line. The mixed line represents a simple approximation of the relative
    RMSE for the subset simulation algorithm: $\frac{T_\alpha}{m}\, \frac{1 - p_0}{p_0}$,
    where $T_\alpha = \lceil \frac{\log \alpha}{\log p_0} \rceil$.
  }
  \label{fig:RMSE}
\end{figure}

\begin{figure}
  \centering
  
  \subfloat[Example~\ref{ex:FB}: Four branch]{%
    \includegraphics{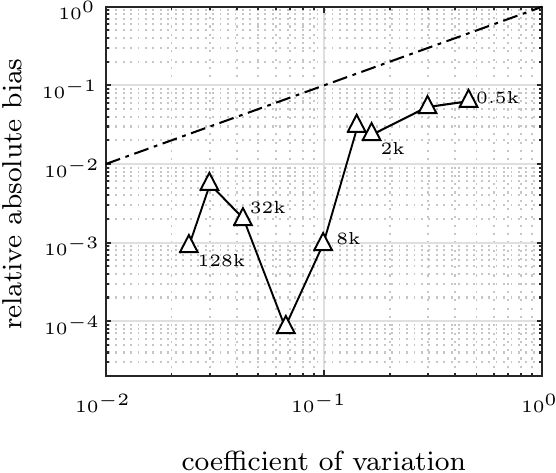}}%
  \hspace{5mm}
  \subfloat[Example~\ref{ex:DCB}: Cantilever beam]{%
    \includegraphics{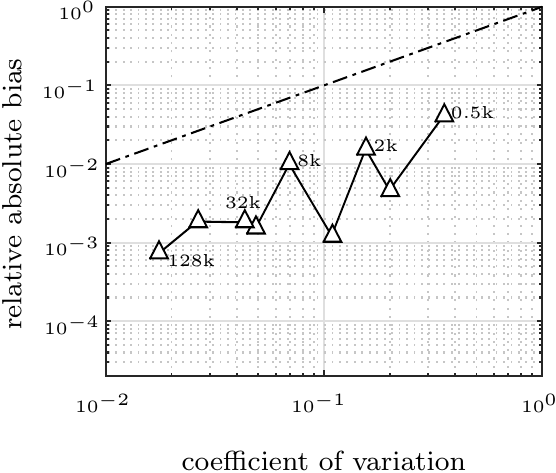}}%

  \smallskip 

  \subfloat[Example~\ref{ex:RNO}: Nonlinear oscillator]{%
    \includegraphics{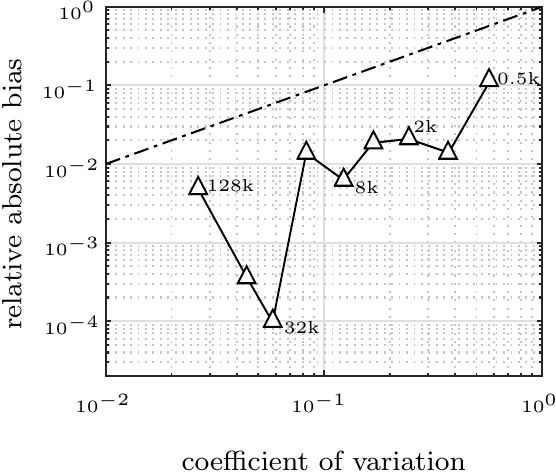}}%

  \smallskip

  \caption{Relative absolute bias of the BSS estimator as a function of its
    coefficient of variation, estimated using 100 independent runs. The relative
    absolute bias is estimated using, for each test case, the reference
    value~$\alpha_{\text{ref}}$ provided in Table~\ref{tab:summary-test-cases}.}
  \label{fig:rab-vs-cov}
\end{figure}

Figure~\ref{fig:RMSE} represents the relative Root-Mean-Square Error (RMSE) of
all three algorithms, as a function of the average number of evaluations. For
the subset simulation algorithm, the number of evaluations is directly
proportional to~$m$ and the RMSE decreases as expected like~$\frac{1}{m}$ (with
a constant much smaller than that of plain Monte Carlo simulation). \TwoSMART
clearly outperforms subset simulation, but offers no simple way of tuning the
accuracy of the final estimate (which is why only one result is presented, using
the default settings of the algorithm). Finally, BSS clearly and consistently
outperforms both \TwoSMART and subset simulation on these three examples: the
relative RMSE goes to zero at a rate much faster than subset simulation's (a
feature that is made possible by the smoothness of the limit-state function,
which is leveraged by the Gaussian process model), and the sample size~$m$ is
the only tuning parameters that needs to be acted upon in order change the
accuracy of the final estimate. Figure~\ref{fig:rab-vs-cov} provides more
insight into the error of the BSS estimate by confirming that, as intended by
design of the adaptive stopping rule, variance is the main component of the RMSE
(in other words, the bias is negligible in these examples).

Finally, note that the BSS estimation involves a computational overhead with
respect to subset simulation.  A careful analysis of the run times of BSS on the
three examples (provided as Supplementary Material~\ref{SM:runtime:overhead})
reveals that the computational overhead of BSS is approximately equal to
$C_0 + C_1\, m\, N_{\mathrm{SUR}}$, where $N_{\mathrm{SUR}}$ the total number of
evaluations selected using the SUR criterion (i.e., all evaluations except the
initial design of experiments).  This shows that, in our implementation, the
most time-consuming part of the algorithm is the selection of additional
evaluation points using the SUR criterion.  However, in spite of its
computational overhead, BSS is preferable to the subset simulation algorithm in
terms of computation time, on the three test cases, as soon as the evaluation
time~$\tau_{\text{sim}}$ of~$f$ is large enough---larger than, say, 10~ms for
the considered range of relative RMSE (see Supplementary
Material~\ref{SM:runtime:extrapolation} for details).  Consider for instance
Example~\ref{ex:FB} with~$\tau_{\mathrm{sim}} = 1\, s$: BSS with sample
size~$m = 8000$ achieves a relative RMSE of approximately~$10\%$ in about
3~minutes\footnote{This computation time can be further decomposed as follows:
  1~minute of evaluation time, corresponding to $\bar N = 63.2$ evaluations on
  average (see Figure~\ref{fig:nbEvalsHisto}) and 2~minutes of algorithm
  overhead.}  while subset simulation requires about 19~hours to achieve a
comparable accuracy.


\section{Discussion}
\label{sec:discuss}

We propose an algorithm called Bayesian subset simulation for the estimation of
small probabilities of failure---or more generally the estimation of the volume
of excursion of a function above a threshold---when the limit-state function is
expensive to evaluate. This new algorithm is built upon two key techniques: the
SMC method known as subset simulation or adaptive multilevel splitting on the
one hand, and the Bayesian (Gaussian process based) SUR sampling strategy on the
other hand. SMC simulation provides the means for evaluating the Bayesian
estimate of the probability of failure, and to evaluate and optimize the SUR
sampling criterion. In turn, the SUR sampling strategy makes it possible to
estimate the level sets of the (smooth) limit-state function using a restricted
number of evaluations, and thus to build a good sequence of target density for
SMC simulation. Our numerical experiments show that this combination achieves
significant savings in evaluations on three classical examples from the
structural reliability literature.

An adaptive stopping rule is used in the BSS algorithm to choose the number of
evaluation added by the SUR sampling strategy at each stage. Evaluations at
intermediate stages are not directly useful to refine the final probability
estimate, but their importance must not be overlooked: they make it possible to
learn in a robust way the level sets of the limit-state function, and therefore
to build a sequence of densities that converges to the boundary of the failure
region. Achieving a better understanding of the connection between the number of
evaluations spent on intermediate level sets and the robustness of the algorithm
is an important perspective for future work. In practice, if the budget of
evaluations permits, we recommend running several passes of the BSS algorithm,
with decreasing tolerances for the adaptive stopping rule, to make sure that no
failure mode has been missed.

The adaptive stopping rule also makes it possible to refine the estimation of
the final level set to make sure that the posterior model is good enough with
respect to the SMC sample size.  Other settings of the stopping rule could of
course be considered.  For instance, BSS could stop when the bias is expected to
be of the same order as the standard deviation. Future work will focus on fully
automated variants on the BSS algorithm, where the number of evaluations and the
SMC sample size would be controlled in order to achieve a prescribed error
level.

\appendix

\section{Computation of the variance in the idealized setting}
\label{app:variance}

This section provides a derivation of Equations~\eqref{eq:var-bss-1}
and~\eqref{eq:var-bss-2}, together with an explicit expression of the estimated
coefficient of variation~$\hat\delta_{m, T}$ used in
Section~\ref{sec:bss-settings}. Both are obtained in the setting of the
idealized BSS algorithm described in Section~\ref{subsec:adapt-thresh}, where
the samples~$\YY{t} = \left\{ Y_t^1,\, \ldots,\, Y_t^m \right\}$ are assumed
\iid (with density~$q_t$) and mutually independent.

Recall from Equation~\eqref{eq:bss-estimator} that the BSS estimator can be
written as
\begin{equation}
  \label{eq:bss-estimator-recall}
  \alphaBSS_T
  \;=\; %
  \prod_{t=1}^T \frac{\alphaBSS_t}{\alphaBSS_{t - 1}}
  \;=\; %
  \prod_{t=1}^T \left(
    \frac{1}{m}\, \sum_{j=1}^{m}
    \frac{g_t(\Y{t-1}{j})}{g_{t-1}(\Y{t-1}{j})}
  \right) %
  \;=\; %
  \prod_{t=1}^T\, \pBSS_t\,,
\end{equation}
where we have set, for all $t \in \{ 1,\, \ldots,\, T \}$,
\begin{equation*}
  \pBSS_t = \frac{1}{m}\, \sum_{j=1}^{m}
  \frac{g_t(\Y{t-1}{j})}{g_{t-1}(\Y{t-1}{j})}.
\end{equation*}
Observe that the random variables~$\pBSS_t$ are independent, with mean
\begin{equation*}
  \EE\left( \pBSS_t \right) %
  = \int_\XX \frac{g_t}{g_{t-1}}\, q_{t-1} %
  = \int_\XX \frac{g_t}{g_{t-1}}\, \frac{g_{t-1}\, \ddpX}{\alphaB_{t-1}} %
  = \frac{\alphaB_t}{\alphaB_{t - 1}} %
  = \pB_t
\end{equation*}
and variance
\begin{align*}
  \var\left( \pBSS_t \right) %
  & = \frac{1}{m}\, \var\left( %
    \frac{g_t\left( Y_{t-1}^1 \right)}{g_{t-1}\left( Y_{t-1}^1 \right)}
  \right) \\
  & = \frac{1}{m}\, \left[ %
    \int_\XX \frac{g_t^2}{g_{t-1}^2}\, \frac{g_{t-1}\, \ddpX}{\alphaB_{t-1}}
    - \left( \frac{\alphaB_t}{\alphaB_{t - 1}} \right)^2
  \right] \\
  & =  \frac{1}{m}\, \left[ %
    \frac{1}{\alphaB_{t-1}}\, \int_\XX \frac{g_t^2}{g_{t-1}}\, \ddpX
    - \left( \frac{\alphaB_t}{\alphaB_{t - 1}} \right)^2
  \right] \\
  & = \frac{1}{m}\, \left( \pB_t \right)^2 \kappa_t,
\end{align*}
where $\kappa_t$ is defined by~\eqref{eq:var-bss-2}. Therefore, the coefficients
of variation~$\delta_{m, t}$ of the sequence of estimators~$\alphaBSS_t$ obey
the recurrence relation $\delta_{m, t}^2 = \frac{1}{m} \kappa_t %
  + \left( 1 + \frac{1}{m} \kappa_t \right)\, \delta_{m, t - 1}^2$,
and we conclude that
\begin{align*}
  \delta_{m, T}^2 & = \frac{1}{m} \kappa_T %
  + \left( 1 + \frac{1}{m} \kappa_T \right)\, \frac{1}{m} \kappa_{T - 1} \\
  & \quad + \left( 1 + \frac{1}{m} \kappa_T \right)\, 
  \left( 1 + \frac{1}{m} \kappa_{T - 1} \right)\, \frac{1}{m} \kappa_{T - 2} %
  + \cdots \\
  & = \frac{1}{m} \sum_{t = 1}^T \kappa_t + O\left( \frac{1}{m^2} \right),
\end{align*}
which proves Equations~\eqref{eq:var-bss-1}--\eqref{eq:var-bss-2}. We construct
an estimator of the coefficient of variation recursively, using the relation
\begin{equation*}
  \hat\delta_{m, t}^2 = \frac{1}{m} \hat\kappa_t %
  + \left( 1 + \frac{1}{m} \hat\kappa_t \right)\, \hat\delta_{m, t - 1}^2,
\end{equation*}
with
\begin{equation*}
  \hat\kappa_t = %
  \left( \pBSS_t \right)^{-2} \,\cdot\, %
  \frac{1}{m} \sum_{j=1}^m \left( %
    \frac{g_t\left( Y_{t-1}^j \right)}{g_{t-1}\left( Y_{t-1}^j \right)}
    - \pBSS_t
  \right)^2.
\end{equation*}

\bibliographystyle{siamplain}
\bibliography{bss-paper}

\begin{thebibliography}{1}

\bibitem{chang2003optim}
{\sc B.~L. Chang, A.~Doucet, and V.~B. Tadic}, {\em Optimisation of particle
  filters using simultaneous perturbation stochastic approximation}, in
  Proceedings of the 2003 IEEE International Conference on Acoustics, Speech
  and Signal Processing (ICASSP), 2003, pp.~681--684.

\bibitem{chevalier:SM}
{\sc C.~Chevalier, J.~Bect, D.~Ginsbourger, E.~Vazquez, V.~Picheny, and
  Y.~Richet}, {\em Fast parallel kriging-based stepwise uncertainty reduction
  with application to the identification of an excursion set}, Technometrics,
  56 (2013), pp.~455--465.

\bibitem{cornebise2008}
{\sc J.~Cornebise, {\'E}.~Moulines, and J.~Olsson}, {\em Adaptive methods for
  sequential importance sampling with application to state space models}, Stat.
  Comput., 18 (2008), pp.~461--480.

\bibitem{fearnhead2013adaptive}
{\sc P.~Fearnhead and B.~M. Taylor}, {\em An adaptive sequential monte carlo
  sampler}, Bayesian analysis, 8 (2013), pp.~411--438.

\bibitem{rasmussenWilliams}
{\sc C.~E. Rasmussen and C.~K.~I. Williams}, {\em Gaussian Processes for
  Machine Learning}, MIT Press, 2006.

\bibitem{roberts1997weak}
{\sc G.~O. Roberts, A.~Gelman, and W.~R. Gilks}, {\em Weak convergence and
  optimal scaling of random walk {M}etropolis algorithms}, The Annals of
  Applied Probability, 7 (1997), pp.~110--120.

\bibitem{scheuerer2010regularity}
{\sc M.~Scheuerer}, {\em Regularity of the sample paths of a general second
  order random field}, Stochastic Processes and their Applications, 120 (2010),
  pp.~1879--1897.

\bibitem{Ste99:SM}
{\sc M.~L. Stein}, {\em Interpolation of Spatial Data: Some Theory for
  {K}riging}, Springer, New York, 1999.

\bibitem{vihola2010convergence}
{\sc M.~Vihola}, {\em On the convergence of unconstrained adaptive {M}arkov
  chain {M}onte {C}arlo algorithms}, PhD thesis, University of Jyväskylä, 2010.

\end{thebibliography}


\begin{thebibliography}{10}

\bibitem{arnaud:2010:quant}
{\sc A.~Arnaud, J.~Bect, M.~Couplet, A.~Pasanisi, and E.~Vazquez}, {\em
  {{\'E}valuation d'un risque d'inondation fluviale par planification
  s{\'e}quentielle d'exp{\'e}riences}}, in 42èmes Journées de Statistique (JdS
  2010), Marseille, France, May 24--28, 2010.

\bibitem{au01:_estim}
{\sc S.~K. Au and J.~Beck}, {\em Estimation of small failure probabilities in
  high dimensions by subset simulation}, Probab. Eng. Mech., 16 (2001),
  pp.~263--277.

\bibitem{auffray2014rare}
{\sc Y.~Auffray, P.~Barbillon, and J.-M. Marin}, {\em Bounding rare event
  probabilities in computer experiments}, Comput. Statist. Data Anal., 80
  (2014), pp.~153--166.

\bibitem{Balesdent2013}
{\sc M.~Balesdent, J.~Morio, and J.~Marzat}, {\em {K}riging-based adaptive
  {I}mportance {S}ampling algorithms for rare event estimation}, Struct. Saf.,
  44 (2013), pp.~1--10.

\bibitem{bayarri:2009:volcanic}
{\sc M.~J. Bayarri, J.~O. Berger, E.~S. Calder, K.~Dalbey, S.~Lunagomez, A.~K.
  Patra, E.~B. Pitman, E.~T. Spiller, and R.~L. Wolpert}, {\em Using
  statistical and computer models to quantify volcanic hazards}, Technometrics,
  51 (2009), pp.~402--413.

\bibitem{bect:2012:stco}
{\sc J.~Bect, D.~Ginsbourger, L.~Li, V.~Picheny, and E.~Vazquez}, {\em
  Sequential design of computer experiments for the estimation of a probability
  of failure}, Stat. Comput., 22 (2012), pp.~773--793.

\bibitem{stktoolbox}
{\sc J.~Bect, E.~Vazquez, et~al.}, {\em {STK}: a {S}mall ({M}atlab/{O}ctave)
  {T}oolbox for {K}riging. {R}elease 2.4 (to appear)}, 2016,
  \url{http://kriging.sourceforge.net}.

\bibitem{bourinet2010}
{\sc J.-M. Bourinet}, {\em Ferum 4.1 user's guide}.
\newblock \url{http://www.ifma.fr/FERUM}, 2010.

\bibitem{Bourinet2016}
{\sc J.-M. Bourinet}, {\em Rare-event probability estimation with adaptive
  support vector regression surrogates}, Reliab. Eng. Syst. Saf., 150 (2016),
  pp.~210--221.

\bibitem{bourinet2011assessing}
{\sc J.-M. Bourinet, F.~Deheeger, and M.~Lemaire}, {\em Assessing small failure
  probabilities by combined subset simulation and support vector machines},
  Struct. Saf., 33 (2011), pp.~343--353.

\bibitem{brehier2016}
{\sc C.-E. Bréhier, L.~Goudenège, and L.~Tudela}, {\em Central limit theorem
  for adaptive multilevel splitting estimators in an idealized setting}, in
  Monte Carlo and Quasi-Monte Carlo Methods (MCQMC 2014), R.~Cools and
  D.~Nuyens, eds., Springer, 2016.

\bibitem{Cadini2014}
{\sc F.~Cadini, F.~Santos, and E.~Zio}, {\em An improved adaptive kriging-based
  importance technique for sampling multiple failure regions of low
  probability}, Reliab. Eng. Syst. Saf., 131 (2014), pp.~109--117.

\bibitem{cannamela:2008}
{\sc C.~Cannamela, J.~Garnier, and B.~Iooss}, {\em Controlled stratification
  for quantile estimation}, Annals of Applied Statistics, 2 (2008),
  pp.~1554--1580.

\bibitem{chevalier:moda10}
{\sc C.~Chevalier, J.~Bect, D.~Ginsbourger, and I.~Molchanov}, {\em Estimating
  and quantifying uncertainties on level sets using the {V}orob'ev expectation
  and deviation with gaussian process models}, in mODa 10 --- Advances in
  Model-Oriented Design and Analysis, Contributions to Statistics, Springer,
  2013, pp.~35--43.

\bibitem{chevalier}
{\sc C.~Chevalier, J.~Bect, D.~Ginsbourger, E.~Vazquez, V.~Picheny, and
  Y.~Richet}, {\em Fast parallel kriging-based stepwise uncertainty reduction
  with application to the identification of an excursion set}, Technometrics,
  56 (2013), pp.~455--465.

\bibitem{chopin:2002}
{\sc N.~Chopin}, {\em A sequential particle filter method for static models},
  Biometrika, 89 (2002), pp.~539--552.

\bibitem{cerou2012}
{\sc F.~Cérou, P.~Del~Moral, T.~Furon, and A.~Guyader}, {\em Sequential {M}onte
  {C}arlo for rare event estimation}, Stat. Comput., 22 (2012), pp.~795--808.

\bibitem{deroc:2008:chap5}
{\sc E.~De~Rocquigny, N.~Devictor, S.~Tarantola, et~al.}, {\em Determination of
  the risk due to personal electronic devices (PEDs) carried out on
  radio-navigation systems aboard aircraft}, in  \cite{deroc:2008:book}, 2008,
  ch.~5, pp.~65--80.

\bibitem{deroc:2008:book}
{\sc E.~De~Rocquigny, N.~Devictor, S.~Tarantola, et~al.}, {\em Uncertainty in
  industrial practice}, Wiley, 2008.

\bibitem{deheeger:2008}
{\sc F.~Deheeger}, {\em Couplage mécano-fiabiliste: ${}^2${SMART} --
  Méthodologie d'appren\-tissage stochastique en fiabilité}, PhD thesis,
  Université B. Pascal (Clermont-Ferrand~II), 2008.

\bibitem{delmoral:2006:sequential}
{\sc P.~Del~Moral, A.~Doucet, and A.~Jasra}, {\em Sequential {M}onte {C}arlo
  samplers}, J. the Royal Statistical Society: Series B (Statistical
  Methodology), 68 (2006), pp.~411--436.

\bibitem{DelMoral2012}
{\sc P.~Del~Moral, A.~Doucet, and A.~Jasra}, {\em An adaptive sequential
  {M}onte {C}arlo method for approximate {B}ayesian computation}, Stat.
  Comput., 22 (2012), pp.~1009--1020.

\bibitem{diaconis:1995:three}
{\sc P.~Diaconis and S.~Holmes}, {\em Three examples of Monte-Carlo Markov
  chains: at the interface between statistical computing, computer science, and
  statistical mechanics}, vol.~72 of IMA volumes in Mathematics and its
  Applications, Springer, 1995, pp.~43--56.

\bibitem{douc2005comparison}
{\sc R.~Douc and O.~Capp{\'{e}}}, {\em Comparison of resampling schemes for
  particle filtering}, in Proceedings of the 4th International Symposium on
  Image and Signal Processing and Analysis (ISPA), 2005, pp.~64--69.

\bibitem{douc:2008:limit}
{\sc R.~Douc and E.~Moulines}, {\em Limit theorems for weighted samples with
  applications to sequential {M}onte {C}arlo methods}, The Annals of
  Statistics, 36 (2008), pp.~2344--2376.

\bibitem{Dubourg:2011:phd}
{\sc V.~Dubourg}, {\em Adaptive surrogate models for reliability analysis and
  reliability-based design optimization}, PhD thesis, Universit\'e Blaise
  Pascal -- Clermont II, 2011.

\bibitem{dubourg:icasp11}
{\sc V.~Dubourg, F.~Deheeger, and B.~Sudret}, {\em Metamodel-based importance
  sampling for the simulation of rare events}, in 11th International Conference
  on Applications of Statistics and Probability in Civil Engineering (ICASP
  11), 2011.

\bibitem{dubourg:mbis}
{\sc V.~Dubourg, F.~Deheeger, and B.~Sudret}, {\em Metamodel-based importance
  sampling for structural reliability analysis}, Probab. Eng. Mech., 33 (2013),
  pp.~47--57.

\bibitem{dubourg:2011:rbdo}
{\sc V.~Dubourg, B.~Sudret, and J.-M. Bourinet}, {\em Reliability-based design
  optimization using kriging surrogates and subset simulation}, Struct.
  Multiscip. Optim., 44 (2011), pp.~673--690.

\bibitem{echard:2011:akmcs}
{\sc B.~Echard, N.~Gayton, and M.~Lemaire}, {\em {AK-MCS}: An active learning
  reliability method combining kriging and {M}onte {C}arlo simulation}, Struct.
  Saf., 33 (2011), pp.~145--154.

\bibitem{Echard:2013}
{\sc B.~Echard, N.~Gayton, M.~Lemaire, and N.~Relun}, {\em A combined
  importance sampling and kriging reliability method for small failure
  probabilities with time-demanding numerical models}, Reliab. Eng. Syst. Saf.,
  111 (2013), pp.~232--240.

\bibitem{feliot16:_bayes}
{\sc P.~Feliot, J.~Bect, and V.~E.}, {\em A {B}ayesian approach to constrained
  single- and multi-objective optimization}, J. Global Optim.,  (in press),
  pp.~1--37.

\bibitem{lgplv21}
{\sc {Free~Software~Foundation}}, {\em {GNU} {L}esser {G}eneral {P}ublic
  {L}icense, version~2.1},
  \url{http://www.gnu.org/licenses/old-licenses/lgpl-2.1.html}.

\bibitem{garcia:2010:emc}
{\sc E.~Garcia}, {\em Electromagnetic compatibility uncertainty, risk, and
  margin management}, IEEE Trans. Electromag. Compat., 52 (2010), pp.~3--10.

\bibitem{gayton2003cq}
{\sc N.~Gayton, J.~M. Bourinet, and M.~Lemaire}, {\em {CQ2RS}: a new
  statistical approach to the response surface method for reliability
  analysis}, Struct. Saf., 25 (2003), pp.~99--121.

\bibitem{glasserman:1999:mls}
{\sc P.~Glasserman, P.~Heidelberger, P.~Shahabuddin, and T.~Zajic}, {\em
  Multilevel splitting for estimating rare event probabilities}, Oper. Res., 47
  (1999), pp.~585--600.

\bibitem{Huang2016}
{\sc X.~Huang, J.~Chen, and H.~Zhu}, {\em Assessing small failure probabilities
  by {AK-SS}: An active learning method combining kriging and subset
  simulation}, Struct. Saf., 59 (2016), pp.~86--95.

\bibitem{JMY90}
{\sc M.~E. Johnson, L.~M. Moore, and D.~Ylvisaker}, {\em Minimax and maximin
  distance designs}, J. Statist. Plan. Inference, 26 (1990), pp.~131--148.

\bibitem{jonkman:2008:flood}
{\sc S.~N. Jonkman, M.~Kok, and J.~K. Vrijling}, {\em Flood risk assessment in
  the netherlands: A case study for dike ring south holland}, Risk Analysis, 28
  (2008), pp.~1357--1374.

\bibitem{lily:phd}
{\sc L.~Li}, {\em Sequential Design of Experiments to Estimate a Probability of
  Failure}, PhD thesis, {Sup{\'e}lec}, May 2012,
  \url{https://tel.archives-ouvertes.fr/tel-00765457}.

\bibitem{lilyPSAM}
{\sc L.~Li, J.~Bect, and E.~Vazquez}, {\em Bayesian subset simulation: a
  kriging-based subset simulation algorithm for the estimation of small
  probabilities of failure}, in Proceedings of PSAM 11 \& ESREL 2012, 25-29
  June 2012, Helsinki, Finland., 2012.

\bibitem{liu:2008:book}
{\sc J.~S. Liu}, {\em {M}onte {C}arlo strategies in scientific computing},
  Springer, 2008.

\bibitem{loeppky2009css}
{\sc J.~L. Loeppky, J.~Sacks, and W.~J. Welch}, {\em Choosing the sample size
  of a computer experiment: A practical guide}, Technometrics, 51 (2009),
  pp.~366--376.

\bibitem{melchers:1999:book}
{\sc R.~E. Melchers}, {\em Structural Reliability: Analysis and Prediction.
  Second Edition.}, Wiley, 1999.

\bibitem{MM95}
{\sc M.~D. Morris and T.~J. Mitchell}, {\em Exploratory designs for
  computational experiments}, J. Statist. Plan. Inference, 43 (1995),
  pp.~381--402.

\bibitem{oakley:2004:perc}
{\sc J.~Oakley}, {\em Estimating percentiles of uncertain computer code
  outputs}, J. Roy. Statist. Soc. Ser. C, 53 (2004), pp.~83--93.

\bibitem{oconnor:2012:book}
{\sc P.~P. O'Connor and A.~Kleyner}, {\em Practical Reliability Engineering},
  Wiley, 2012.

\bibitem{rausand:2004:book}
{\sc M.~Rausand and A.~Hoyland}, {\em System reliability theory: models and
  statistical methods (second edition)}, Wiley, 2004.

\bibitem{robert:2004:monte}
{\sc C.~P. Robert and G.~Casella}, {\em {M}onte {C}arlo statistical methods,
  2nd edition}, Springer Verlag, 2004.

\bibitem{santner:2003:dace}
{\sc T.~J. Santner, B.~J. Williams, and W.~Notz}, {\em {T}he {D}esign and
  {A}nalysis of {C}omputer {E}xperiments}, Springer Verlag, 2003.

\bibitem{Ste99}
{\sc M.~L. Stein}, {\em Interpolation of Spatial Data: Some Theory for
  {K}riging}, Springer, New York, 1999.

\bibitem{vaz:09:sysid}
{\sc E.~Vazquez and J.~Bect}, {\em A sequential {B}ayesian algorithm to
  estimate a probability of failure}, in Proceedings of the 15th {IFAC}
  {S}ymposium on {S}ystem {I}dentification, {SYSID} 2009, {S}aint-{M}alo
  {F}rance, 2009.

\bibitem{villemonteix:2009:iago}
{\sc J.~Villemonteix, E.~Vazquez, and E.~Walter}, {\em An informational
  approach to the global optimization of expensive-to-evaluate functions}, J.
  Global Optim., 44 (2009), pp.~509--534.

\bibitem{waarts:2000:phd}
{\sc P.~H. Waarts}, {\em Structural reliability using finite element methods},
  PhD thesis, Delft University of Technology, 2000.

\bibitem{zio:2009:loca}
{\sc E.~Zio and N.~Pedroni}, {\em Estimation of the functional failure
  probability of a thermal-hydraulic passive system by subset simulation},
  Nuclear Eng. Design, 239 (2009), pp.~580--599.

\bibitem{Zuev2012}
{\sc K.~M. Zuev, J.~L. Beck, S.-K. Au, and L.~S. Katafygiotis}, {\em Bayesian
  post-processor and other enhancements of subset simulation for estimating
  failure probabilities in high dimensions}, Computers \& Structures, 92--93
  (2012), pp.~283--296.

\end{thebibliography}

\pagebreak

\renewcommand\rightmark{%
  BAYESIAN SUBSET SIMULATION (SUPPLEMENTARY MATERIALS)}

\makeatletter
\renewcommand\section{\@startsection{section}{1}{.25in}%
  {1.3ex \@plus .5ex \@minus .2ex}%
  {-.5em \@plus -.1em}%
  {\reset@font\color{header1}\normalsize\HLtext}}
\makeatother

\setcounter{equation}{0}  \renewcommand\theequation {SM\arabic{equation}}
\setcounter{figure}{0}    \renewcommand\thefigure   {SM\arabic{figure}}
\setcounter{table}{0}     \renewcommand\thetable    {SM\arabic{table}}
\setcounter{section}{0}   \renewcommand\thesection  {SM\arabic{section}}
\setcounter{remark}{0}    \renewcommand\theremark   {SM\arabic{remark}}
\setcounter{page}{1}

\renewcommand \bibnumfmt[1]   {[SM#1]}  
\renewcommand \citenumfont[1] {SM#1}    

\begin{center}
  \color{header1}\bfseries\MakeUppercase{Supplementary Materials}
\end{center}

\section{Approximation and optimization of the SUR criterion}
\label{SM:SUR-crit}

This section discusses the numerical procedure that we use for the approximation
and optimization of the SUR criterion used at each stage of the BSS algorithm
(see Sections~\ref{sec:seq-design} and~\ref{sec:implemen}):
\begin{equation*}
  J_n\left( x_{n+1} \right) 
  \;=\;  
  \int_\XX \EE_{n, x_{n+1}} \left( \tau_{n+1,u_t}(x) \right)\, \ddpX(x)\, \dx,
  \qquad n_{t - 1} \le n \le n_t - 1,
\end{equation*}
where we have introduced the simplified notation $\EE_{n, x_{n+1}} \eqdef \EE_n
\left( \,\cdot \bmid X_{n+1} = x_{n+1} \right)$. The numerical approach proposed
here is essentially the same as that used by~\cite{bect:2012:stco}, with a more
accurate way of computing the integrand, following ideas of~\citeSM{chevalier:SM}.

Observing that
\begin{equation*}
  J_n\left( x_{n+1} \right)
  \;=\;
  \alphaB_{t-1}\, \int_\XX \frac{\EE_{n, x_{n+1}} \left( \tau_{n+1,u_t}(x)\, 
      \right)}{g_{t-1}(x)}\, q_{t-1}(x)\, \dx,
\end{equation*}
the integral over~$\XX$ can be approximated, up to a constant, using the
weighted sample~$\YY{t-1}$:
\begin{align}
  J_n\left( x_{n+1} \right)
  & \;\propto\; 
  \int_\XX \frac{\EE_{n, x_{n+1}} \left( \tau_{n+1,u_t}(x) 
    \right)}{g_{t-1}(x)}\, q_{t-1}(x)\, \dx
  \nonumber\\
  & \;\approx\;
  \sum_{j=1}^m w_{t-1}^j\, 
  \frac{\EE_{n, x_{n+1}} \left( \tau_{n+1,u_t}(x)
    \right)_{|x=Y_{t-1}^j}}{g_{t-1}\left( Y_{t-1}^j \right)}.
  \label{eq:SUR-approx-SMC}
\end{align}
Then, simple computations using well-known properties of Gaussian processes
under conditioning allow to obtain an explicit representation of the integrand,
in the spirit of~\citeSM{chevalier:SM}, as a function of the Gaussian process
posterior mean~$\xihat_n$ and posterior covariance~$k_n$:
\begin{align}
  \EE_{n, x_{n+1}} & \left( \tau_{n+1,u_t}(x) \right) \;=\; %
   \Phi \left( \frac{u - \xihat_n(x)}{\sigma_n(x)} \right) %
  + \Phi \left( \frac{u - \xihat_n(x)}{s_n(x, x_{n+1})} \right) 
  \nonumber\\  
  & \qquad - 2\, \Phi_2 \left(
      \begin{pmatrix} u\\ u \end{pmatrix};\,
      \begin{pmatrix} \xihat_n(x)\\ \xihat_n(x) \end{pmatrix},\,
      \begin{pmatrix}
        s_n^2(x, x_{n+1}) & s_n^2(x, x_{n+1})\\
        s_n^2(x, x_{n+1}) & \sigma_n(x)^2
      \end{pmatrix}
      \right),
      \label{eq:SUR-explicit-integrand}
\end{align}
where $\Phi$ is the cumulative distribution function of the normal distribution,
$\Phi_2$ the cumulative distribution function of the bivariate normal
distribution, $\sigma_n^2(x) = k_n(x, x)$ and
$s_n^2(x, x_{n+1}) = k_n(x, x_{n+1})^2 / \sigma_n(x_{n+1})^2$.

The main computational bottleneck, in a direct implementation of the
approximation~\eqref{eq:SUR-approx-SMC} combined with the
representation~\eqref{eq:SUR-explicit-integrand}, is in our experience the
computation of the bivariate cumulative distribution function~$\Phi_2$. Indeed,
assume that the optimization of the approximate criterion is carried out by
means of an exhaustive discrete search on~$\{Y_{t-1}^j, 1 \le j \le m \}$. Then
$m^2$ evaluations of~$\Phi_2$ are required in order select~$X_{n+1}$. To
mitigate this problem, we implemented the pruning idea proposed in Section~3.4
of~\cite{bect:2012:stco}: only a subset of size~$m_0 \le m$ of the set of
particles is actually used, both for the approximation the integral and for the
optimization of the criterion. In this article, the size~$m_0$ is determined
automatically as follows: first, for each particle~$Y_{t-1}^j$, the current
weighted probability of misclassification
\begin{equation*} 
  \widetilde\tau_n^j \eqdef 
  w_{t-1}^j \tau_{n, u_t} \left( Y_{t-1}^j \right) / g_{t-1}\left( Y_{t-1}^j \right) 
\end{equation*}
is computed. Then, the particules are sorted according to the value
of~$\widetilde\tau_n^j$, in decreasing order:
$\widetilde\tau_n^{\varphi(1)} \;\ge\; \widetilde\tau_n^{\varphi(2)} \;\ge\;
\ldots \ge \widetilde\tau_n^{\varphi(m)}$,
and $m_0$ is set to
$\min \left( m_0^{\text{max}}, m_0(\widetilde\tau_n) \right)$, where
$m_0(\widetilde\tau_n)$ is the smallest integer such that
\begin{equation*}
  \sum_{j=1}^{m_0}\widetilde\tau_n^{\varphi(j)}
  \;\ge\; \rho\, \sum_{j=1}^{m}\widetilde\tau_n^j.
\end{equation*}
The values $m_0^{\text{max}} = 1000$ and~$\rho = 0.99$ have been used in all our
simulations.

\section{Stochastic process prior}
\label{SM:spp}

The stochastic process prior used for the numerical experiments in this article
is a rather standard Gaussian process model. We describe it here in full detail
for the sake of completeness. First, $\xi$ is written as
\begin{equation*}
  \xi(x) = \mu + \xi^0(x),
\end{equation*}
where~$\mu \in \RR$ is an unknown constant mean and~$\xi^0$ a zero-mean
stationary Gaussian process with anisotropic covariance function
\begin{equation}
  \label{eq:materncov}
  k(x,y) = \sigma^2 \kappa_{\nu}\left( \sqrt{ \sum_{i=1}^d
      \frac{(x_{[i]} -y_{[i]})^2}{\rho_i^2}}\right)\,, \quad x, y \in \RR^d,
\end{equation}
where $x_{[i]}, y_{[i]}$ denote the $i^{\rm th}$ coordinate of $x$ and $y$, and
$\kappa_\nu$ the Mat\'ern correlation function of regularity~$\nu$ (see
\citeSM{Ste99:SM}, Section~2.10). The scale parameters $\rho_1$, \ldots,
$\rho_d$ (characteristic correlation lengths) are usually called the
\emph{range} parameters of the covariance function.

In this article, the regularity parameter~$\nu$ is set to the fixed value
$\nu = 5/2$, leading to the following analytical expression for the Mat\'ern
correlation function:
\begin{equation}
  \label{equ:matern-corr}
  \kappa_\nu \left( h \right) = %
  \left( 1 + \tilde h + \frac{1}{3} \tilde h^2 \right)\, %
  \exp \left( - \tilde h \right), %
  \qquad \text{with } \tilde h = \sqrt{10}\, \left| h \right|.
\end{equation}
As a consequence, $\xi$ is twice differentiable in the mean-square sense, with
sample paths almost surely in the Sobolev space~$W^{s, 2}$ for all~$s < 5/2$
(\citeSM{scheuerer2010regularity}, Theorem~3).

\begin{remark}
  The parameterization used in Equation~\eqref{equ:matern-corr} is the one
  advocated in~\citeSM{Ste99:SM}. Other parameterizations are sometimes used in
  the literature; e.g., \citeSM{rasmussenWilliams} use
  $\tilde h = \sqrt{5}\, \left| h \right|$.
\end{remark}

\section{Adaptive Metropolis-Hastings algorithm for the move step}
\label{SM:adaptiveMCMC}

A fixed number~$S$ of iterations of a Gaussian Random Walk Metropolis-Hastings
(RWMH) kernel is used for the move step, with adaptation of the standard
deviations of the increments. More precisely, for $s = 1, 2, \ldots, S$,
starting with the set of particles
$\left( Y_{t, 1}^{(j)} \right)_{1 \le j \le m}$ produced by the resampling step,
we first produce perturbed particles:
\begin{equation*}
  \widetilde Y_{t, s}^{(j)} =  Y_{t, s}^{(j)} + %
  \Sigma_{\text{RW}, t, s}\, U_{t, s}, %
  \qquad 1 \le j \le m,
\end{equation*}
where $\Sigma_{\text{RW}, t, s}$ is a diagonal matrix and~$U_{t, s}$ a
$d$-dimensional standard normal vector. The perturbed particle is accepted
as~$Y_{t, s+1}^{(j)}$ with probability
\begin{equation*}
  a_{t,s}^{(j)} = 1 \wedge \frac{q_t\left( \widetilde Y_{t, s}^{(j)} \right)%
  }{q_t\left( Y_{t, s}^{(j)} \right)},
\end{equation*}
and $Y_{t, s}^{(j)}$ is kept otherwise. The~$k^{\text{th}}$ diagonal
element~$\sigma_{\text{RW}, t, s}^{(k)}$ of~$\Sigma_{\text{RW}, t, s}$ is
initialized with
\begin{equation}
  \label{eq:initSigma}
  \sigma_{\text{RW}, 1, 1}^{(k)} = %
  C_{\sigma, \text{init}}\, \sigma_\XX^{(k)},
\end{equation}
where $\sigma_\XX^{(k)}$ is the standard deviation of the~$k^{\text{th}}$
marginal of~$\PX$, and then updated using
\begin{equation}
  \label{eq:updateSigma}
  \log \sigma_{\text{RW}, t, s}^{(k)} = %
  \begin{cases}
    \log \sigma_{\text{RW}, t, s - 1}^{(k)} %
    + \Delta_\sigma / s %
    & \text{if } \bar a_{t, s} > a_{\text{target}} \\
    \log \sigma_{\text{RW}, t, s - 1}^{(k)} %
    - \Delta_\sigma / s %
    & \text{otherwise,}
  \end{cases}
\end{equation}
where $\sigma_{\text{RW}, t, -1}^{(k)} = \sigma_{\text{RW}, t-1, S}^{(k)}$,
$\bar a_{t, s} = \frac{1}{m} \sum_{j=1}^m a_{t,s}^{(j)}$ is the average
acceptance probability and $a_{\text{target}}$ some prescribed target value.

The following parameter values have been used for the numerical simulations
presented in this article: $S = 10$, $C_{\sigma, \text{init}} = 2 / \sqrt{d}$,
$\Delta_\sigma = \log(2)$ and $a_{\text{target}} = 30\%$.

\begin{remark}
  The $1/\sqrt{d}$ scaling for the constant~$C_{\sigma, \text{init}}$
  in~\eqref{eq:initSigma} is motivated by the well-known theorem of
  \citeSM{roberts1997weak}, which provides the optimal covariance
  matrix~$2.38^2 \Sigma / d$ for a Gaussian target with covariance
  matrix~$\Sigma$ in high dimension.
\end{remark}

\begin{remark}
  The value $\Delta_\sigma = \log(2)$ that was used in the simulation
  turns out to be too small to be a good general recommendation for a
  default value. %
  Indeed, with $S = 10$ steps of adaptation, it leads to a maximal
  adaptation factor
  of~$2^{1 + \frac{1}{2} + \ldots + \frac{1}{10}} \approx 7.6$, which
  might prove too small for some cases. %
  The value $\Delta_\sigma = \log(10)$ is thus used as a default value
  in the Matlab/Octave program that is provided as Supplementary
  Material.
\end{remark}

\begin{remark}
  The reader is referred to \citeSM{vihola2010convergence} and references
  therein for more information on adaptive MCMC algorithms, including adaptive
  scaling algorithms such as~\eqref{eq:updateSigma}. Note, however, that the
  adaptation scheme proposed here is not strictly-speaking an adaptive MCMC
  scheme, since we use the entire population of particles to estimate the
  acceptance probability.
\end{remark}

\begin{remark}
  Our adaptive scheme~\eqref{eq:initSigma}--\eqref{eq:updateSigma} is admittedly
  very simple, but works well in the examples of the paper.  More refined
  adaptation schemes might be needed to deal with harder problems.  For
  instance, it might become necessary to implement a truly anisotropic
  adaptation of the covariance matrix, or to use different kernels in different
  regions of the input space.  Ideas from the adaptive SMC literature (see,
  e.g., \citeSM{fearnhead2013adaptive, cornebise2008, chang2003optim}) could be
  leveraged to achieve these goals, which fall out of the scope of the present
  paper.
\end{remark}

\pagebreak  

\section{Run time of the BSS algorithm}
\label{SM:runtime}

\subsection{Overhead of the BSS algorithm}
\label{SM:runtime:overhead}
The median run times of the Bayesian subset simulation (BSS) and subset
simulation (SS) algorithms on the three test cases of Section~\ref{sec:num} are
reported in Table~\ref{SM:tab:runtimes}.  Since the function~$f$ in each of
these examples is actually very fast to evaluate, these run times provide a good
indication of the numerical complexity of the algorithms. Consequently, they
provide a measure of the \emph{overhead} of the algorithm if it were actually
run on an expensive-to-evaluate function (i.e., the fraction of the computation
time not dedicated to evaluating the function).

As expected, because of the computations related to Gaussian process modeling,
the overhead of BSS is larger than the one of subset simulation, for a given
sample size~$m$.  In our implementation, this overhead, which we denote
by~$\tau_{\text{BSS},0}$, can be explained by a simple linear model:
\begin{equation}
  \label{SM:eq:runtimelinear}
  \tau_{\text{BSS},0} \;\approx\; C_0 + C_1 m N_{\text{SUR}}
\end{equation}
where~$m$ is the sample size and~$N_{\text{SUR}}$ the total number of
evaluations selected using the SUR criterion (i.e., all evaluations except the
initial design of experiments).  On our standard Intel-Nehalem-based
workstation, an ordinary least-square regression using
Equation~\eqref{SM:eq:runtimelinear}, on the $100 \times 3 \times 8 = 2400$
points of our data set (100~runs on 3~cases with 8~sample sizes), yields
$C_0 \approx 27.0$ and~$C_1 \approx 5.2\, 10^{-5}$, with a coefficient of
determination of~$97.2\%$.  We conclude that the most time-consuming part of the
algorithm is the selection of additional evaluation points using the SUR
criterion.

\begin{remark}
  Note that the computation time of BSS would grow as a function
  of~$m^2\, N_{\text{SUR}}$, instead of~$m\, N_{\text{SUR}}$, without the
  pruning idea explained in Section~\ref{app:variance}.
\end{remark}

\subsection{Extrapolation to expensive-to-evaluate functions}
\label{SM:runtime:extrapolation}

\newcommand \rRMSE {\ensuremath{\text{rRMSE}}\xspace}
\newcommand \effBSS {\ensuremath{\varepsilon_{\text{BSS} \,|\, \text{SS}}}\xspace}

\begin{figure}
  \centering
  
  \subfloat[Example~\ref{ex:FB}: Four branch]{%
    \includegraphics{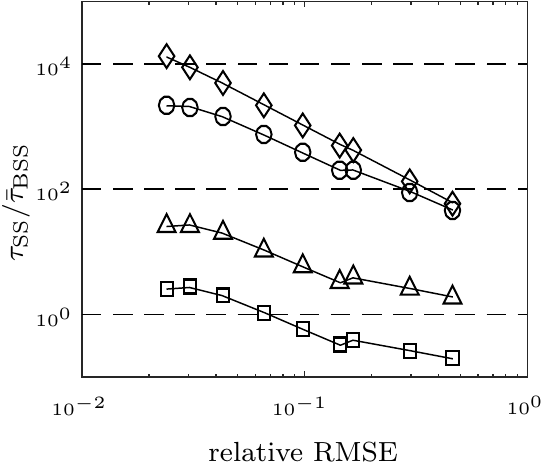}}%
  \hspace{5mm}
  \subfloat[Example~\ref{ex:DCB}: Cantilever beam]{%
    \includegraphics{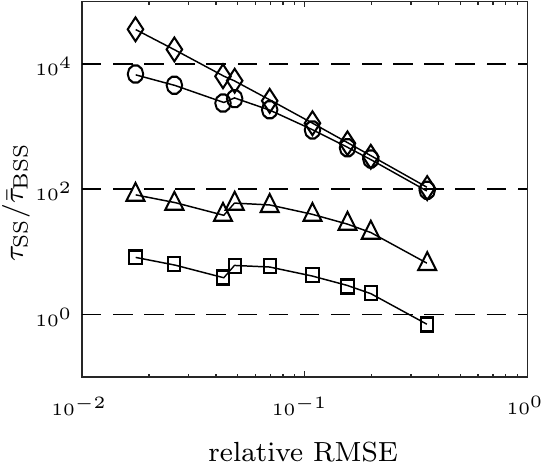}}%

  \smallskip 

  \subfloat[Example~\ref{ex:RNO}: Nonlinear oscillator]{%
    \includegraphics{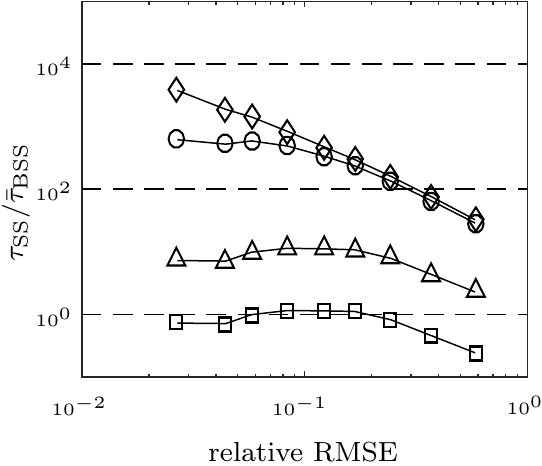}}%

  \smallskip

  \caption{Efficiency of BSS with respect to subset simulation,
    as a function of the relative RMSE, for
    several values of the virtual duration~$\tau_{\text{sim}}$ of a
    single evaluation of the function: $\tau_{\text{sim}} = 1
    \text{ms}$ (squares), $\tau_{\text{sim}} = 10 \text{ms}$ (triangles),
    $\tau_{\text{sim}} = 1 \text{s}$ (circles)
    and $\tau_{\text{sim}} = 1 \text{min}$ (diamonds).}
  \label{SM:fig:runtime}  
\end{figure}

Let us now extrapolate from the available data, obtained for cheap-to-evaluate
test functions, to the case of a non-negligible evaluation time.  To this end,
assume that each evaluation actually costs~$\tau_{\text{sim}}$ in computation
time.  Then, the total run time of BSS becomes
\begin{equation}
  \label{SM:equ:BSSruntime}
  \tau_{\text{BSS}} \;=\; \tau_{\text{BSS},0} + \tau_{\text{sim}}\, N,
\end{equation}
where~$N$ is the total number of evaluations.  For a given test case and a given
sample size~$m$, denote by~$\bar N$ the average total number of evaluation,
$\bar \tau_{\text{BSS},0}$ the average overhead and \rRMSE the relative
root-mean-square error (computed using 100~runs of BSS).  The efficiency of BSS
with respect to the subset simulation algorithm can be measured by the ratio
\begin{equation}
  \label{SM:equ:efficieny}
  \effBSS \;=\;
  \frac{\tau_{\text{SS}}}{\bar \tau_{\text{BSS}}}
\end{equation}
where
$\bar \tau_{\text{BSS}} = \bar \tau_{\text{BSS},0} + \tau_{\text{sim}}\, \bar N$
is the average total run time for BSS, and $\tau_{\text{SS}}$ is an
approximation of the time it would take for subset simulation to reach the same
relative RMSE, computed as:
\begin{equation}
  \label{SM:equ:SSruntime}
  \tau_{\mathrm{SS}} = \tau_{\text{sim}}\, m\,
  \left( 1 + (T_\alpha - 1)(1 - p_0) \right),
\end{equation}
where $T_\alpha = \lceil \frac{\log \alpha}{\log p_0} \rceil$ and $m$ is the
sample size that gives the same relative RMSE according to the relation
\begin{equation}
  \label{SM:equ:approx-SS-RMSE}
  \rRMSE \;\approx\; \frac{T_\alpha}{m}\, \frac{1 - p_0}{p_0}
\end{equation}
(cf.~Figure~\ref{fig:RMSE}). The efficiency~\effBSS is represented on
Figure~\ref{SM:fig:runtime}, for the three test cases, as a function of the
relative RMSE.  It appears clearly for the three examples that, in spite of its
computational overhead due to Gaussian process modeling, BSS is able to provide
a significant time saving as soon as the evaluation time~$\tau_{\text{sim}}$
of~$f$ is large enough---larger than, say, 10~ms for the considered range of
relative RMSE.

\begin{table}
  \centering

  \begin{tabular}{llrrrrrrrrr}
    $m$ &  & 0.5k & 1k & 2k & 4k & 8k & 16k & 32k & 64k & 128k \\
    \hline
    Ex~\ref{ex:FB} 
    & BSS: & 15.9 & 28.3 & 64.7 & 99.2 & 117.8 & 142.6 & 182.1 & 266.5 & 453.8 \\
    & SS:  & 0.1 & 0.1 & 0.1 & 0.1 & 0.2 & 0.3 & 0.7 & 1.3 & 3.1 \\
    \hline
    Ex~\ref{ex:DCB} 
    & BSS: & 3.3 & 3.4 & 4.1 & 5.9 & 10.7 & 20.7 & 41.5 & 70.7 & 120.3 \\
    & SS:  & 0.0 & 0.1 & 0.1 & 0.1 & 0.2 & 0.3 & 0.6 & 1.2 & 2.5 \\
    \hline    
    Ex~\ref{ex:RNO} 
    & BSS: & 5.9 & 7.9 & 10.4 & 16.4 & 30.9 & 64.8 & 156.5 & 378.9 & 994.4 \\
    & SS:  & 0.1 & 0.1 & 0.1 & 0.2 & 0.3 & 0.6 & 1.3 & 2.8 & 6.4 \\
    \hline
  \end{tabular}

  \caption{Run time in seconds of the Bayesian subset simulation (BSS) and subset
    simulation (SS) algorithms on an Intel-Nehalem-based workstation
    (without parallelization).  For  each test case and each sample
    size~$m$,  the reported computation time is the
    median over~$100$ hundred runs of the algorithm.}
  \label{SM:tab:runtimes}

\end{table}

\bibliographystyleSM{siamplain}
\bibliographySM{bss-paper}

\end{document}